\documentclass[conference]{IEEEtran}

\usepackage{amsmath,amssymb,amsfonts,amsthm}
\usepackage{algorithmic}
\usepackage{graphicx}
\usepackage{textcomp}
\usepackage{xcolor}
\usepackage{url}
\usepackage{boxedminipage}
\usepackage{subcaption}
\usepackage{color, colortbl}

\definecolor{lightgraycolor}{gray}{0.85}
\definecolor{darkgraycolor}{gray}{0.5}

\newcommand{\eat}[1] {}
\newcommand{\penmap} {{\tt PenaltyMap}}
\newcommand{\penmapF} {{\tt PenaltyMap-F}}
\newcommand{\LPmap} {{\tt LP-map}}
\newcommand{\LPmapF} {{\tt LP-map-F}}
\renewcommand{\cong} {{\tt cong}}
\newcommand{\dem} {{\tt dem}}
\newcommand{\ncap} {{\tt cap}}
\newcommand{\hatt} {\widehat{t}}
\newcommand{\hatd} {\widehat{d}}
\newcommand{\Spen} {S_{\rm pen}}
\newcommand{\Sopt} {S_{\rm opt}}
\newcommand{\Spi} {S_{\rm \pi}}
\newcommand{\Slp} {S_{\rm LP}}
\newcommand{\VBopt} {V_B^{\rm opt}}
\newcommand{\VBpen} {V_B^{\rm pen}}
\newcommand{\VBpi} {V_B^{\rm \pi}}
\newcommand{\VBlp} {V_B^{\rm LP}}
\newcommand{\LB} {{\rm LB}}
\newcommand{\LP} {{\rm LP}}

\newcommand{\rem} {{\tt rem}}
\newcommand{\calU} {{\cal U}}

\newcommand{\calB} {{\cal B}}
\newcommand{\cost} {{\tt cost}}
\newcommand{\opt} {{\tt opt}}
\newcommand{\wt}[1] {\widetilde{S}}

\newcommand{\havg} {h_{\rm avg}}
\newcommand{\pavg} {p_{\rm avg}}
\newcommand{\pavgstr} {p^*_{\rm avg}}
\newcommand{\bavgstr} {{\tt B}^*_{\rm avg}}
\newcommand{\hmax} {h_{\rm max}}

\newcommand{\mypara}[1] {{\bf #1.}}
\DeclareMathOperator*{\argmin}{argmin}   
\DeclareMathOperator*{\argmax}{argmax}   

\newcommand{\Rightsize} {{\sf Rightsizing}}
\newcommand{\TLrightsize} {{\sf TL\mbox{-}Rightsizing}}
\newcommand{\google} {{GCT-2019}}

\newtheorem{theorem}{\bf Theorem}

\newtheorem{lemma}[theorem]{\bf Lemma}

\begin{document}
\title{Rightsizing Clusters for Time-Limited Tasks}

\author{\IEEEauthorblockN{
Venkatesan T. Chakaravarthy, Padmanabha V. Seshadri, Pooja Aggarwal, Anamitra R. Choudhury\\ 
Ashok Pon Kumar, Yogish Sabharwal, Amith Singhee
}
\IEEEauthorblockA{
\textit{IBM Research - India}\\
\{vechakra, seshapad, aggarwal.pooja, anamchou, ashokponkumar, ysabharwal, asinghee\}@in.ibm.com
}
}

\date{}
\maketitle

\begin{abstract}
Cloud computing has emerged as the dominant platform for application deployment on clusters of computing nodes. In conventional public clouds, designing a suitable initial cluster for a given application workload is important in converging to an appropriate computing foot-print during run-time. In the case of edge or on-premise clouds, cold-start rightsizing the cluster at the time of installation is crucial in avoiding the recurrent capital expenditure. In both these cases, balancing cost-performance trade-off is critical in constructing a cluster of compute nodes for hosting an application with multiple tasks, where each task can demand multiple resources, and the cloud offers nodes with different capacity and cost.

Multidimensional bin-packing algorithms can address this {\em cold-start rightsizing} problem, but these assume that every task 
is always active. In contrast, real-world tasks (e.g. load bursts, batch and dead-lined tasks with time-limits) may be active only during specific time-periods or may have very dynamic load profiles.
The cluster cost can be reduced by reusing resources via time sharing and optimal packing. This motivates our generalized problem of {\em cold-start rightsizing for time-limited tasks}:
given a timeline, time-periods and resource demands for tasks, 
the objective is to place the tasks on a minimum cost cluster of nodes
without violating node capacities at any time instance. 
We design a baseline two-phase algorithm  that performs penalty-based mapping of task to node-type and then, 
solves each node-type independently. 
We prove that the algorithm has an approximation ratio of  $O(D\cdot \min(m, T))$, where $D, m$ and $T$ are the 
number of resources, node-types and timeslots, respectively, 
We then present an improved linear programming based mapping strategy, enhanced further
with a cross-node-type filling mechanism. 
Our experiments on synthetic and real-world cluster traces show significant cost reduction by LP-based mapping compared to the baseline,
and the filling mechanism improves further to produce solutions within $20\%$ of (a lower-bound to) the optimal solution.
\end{abstract}


\section{Introduction}
Cloud computing has emerged as the preferred platform due to its flexibility in planning and provisioning computing clusters with resources for application deployment and scaling. An undersized cluster degrades performance, whereas an oversized cluster causes wastage and drives up cost. Rightsizing addresses this trade-off by constructing an optimal cluster that balances cost and capacity. Rightsizing has been well-studied over a wide spectrum from pre-deployment (cold-start) cluster provisioning to dynamic node provisioning (E.g. \cite{cloud-survey, xu2017cred, yin2016edge, xu2017online, narayanan2017right}). 

Cold-start rightsizing is used as an effective technique to boot-strap an initial cluster based on the workload characteristics, for large public clouds. The capacity of these initial clusters are then managed using dynamic provisioning techniques such as auto-scalers~\cite{autoscaler}. Thus, the cold-start righsizing and dynamic provisioning techniques are complementary in nature. 

Cold-start rightsizing also plays an important role in limited-resource settings where vast resource pools might not exist to support dynamic auto-scalers. In such scenarios, cold-start rightsizing could be the most suitable approach to shape the cluster to a given workload without incurring any capital expenses or logistical challenges. Examples of these scenarios include Telco clouds\cite{5gmec} which co-host computing clusters on base-stations and on-premise clouds dedicated to a single organization. In such infrastructure settings, installation cost on 5G base-stations could be $3\times$~\cite{edgereport} higher than the operational expenditure. Similar challenges exist in on-premise clouds and edge-clouds\cite{kubeedge, ccars, aircraftk8s} which may not have huge resource pools and sufficient number of intra-organization tenants to enable arbitrary on-demand scaling.

This paper deals with cold-start rightsizing that sizes a cluster at the time of installation based on projected demands of the application(s) to be hosted on it. The aim is to address the cost-performance trade-off by constructing an optimal cluster that balances cost and capacity, for a given workload of tasks.

\mypara{Cold-start Rightsizing} 
Consider a workload (set of tasks) consisting of $n$ tasks, where each task demands specific quantities of $D$ resources 
(e.g CPU cores, memory and disk space). The cloud environment offers $m$ types of compute nodes that differ in terms of cost and capacity 
on the $D$ resources, and a replica of a given node-type is a node. Both demands and capacities can be conveniently represented as $D$-dimensional 
vectors. The goal is to purchase a minimum cost cluster of nodes and place each task on a node such that the aggregate demand of 
tasks placed on any node does not exceed its capacity along any dimension (resource). We denote this problem as {\Rightsize}
As discussed below (Prior Work), rightsizing has been well-studied in the contexts of cloud computing, virtual machine migration
and bin packing.

\mypara{Cold-start Rightsizing for Time-Limited Tasks}
While the above formulation targets rightsizing purely based on resource-demands, 
real-world tasks are additionally characterized by timeline properties on 
when they run, how long they run and how much resources they consume at any given time. 

These timeline characteristics are available as direct inputs for batch applications~\cite{springbatch}, for instance, 
batch jobs running at 1:00AM for few hours or a burst in service load during peak hours. Similarly, scheduled tasks with deadlines 
are encountered in edge ~\cite{farris2018federated, varasteh2021holu}. Examples are applications that sample duty-cycled data-sources~\cite{yadav2017self}; or adapting to platform activity (e.g. advanced sleeps modes of 5G New Radio 
which exploit stable user activity patterns~\cite{5gnewradio}).
Since the tasks are active only within a specific time-interval, resources could be reused by other tasks outside the interval. 
This \emph{time-limiting} nature of tasks could be exploited in rightsizing the cluster
leading to cost reduction. With the above motivation, we introduce and study the generalized 
cold-start rightsizing problem for time-limited tasks.

\begin{figure}
\centering
\includegraphics[width=3.5in]{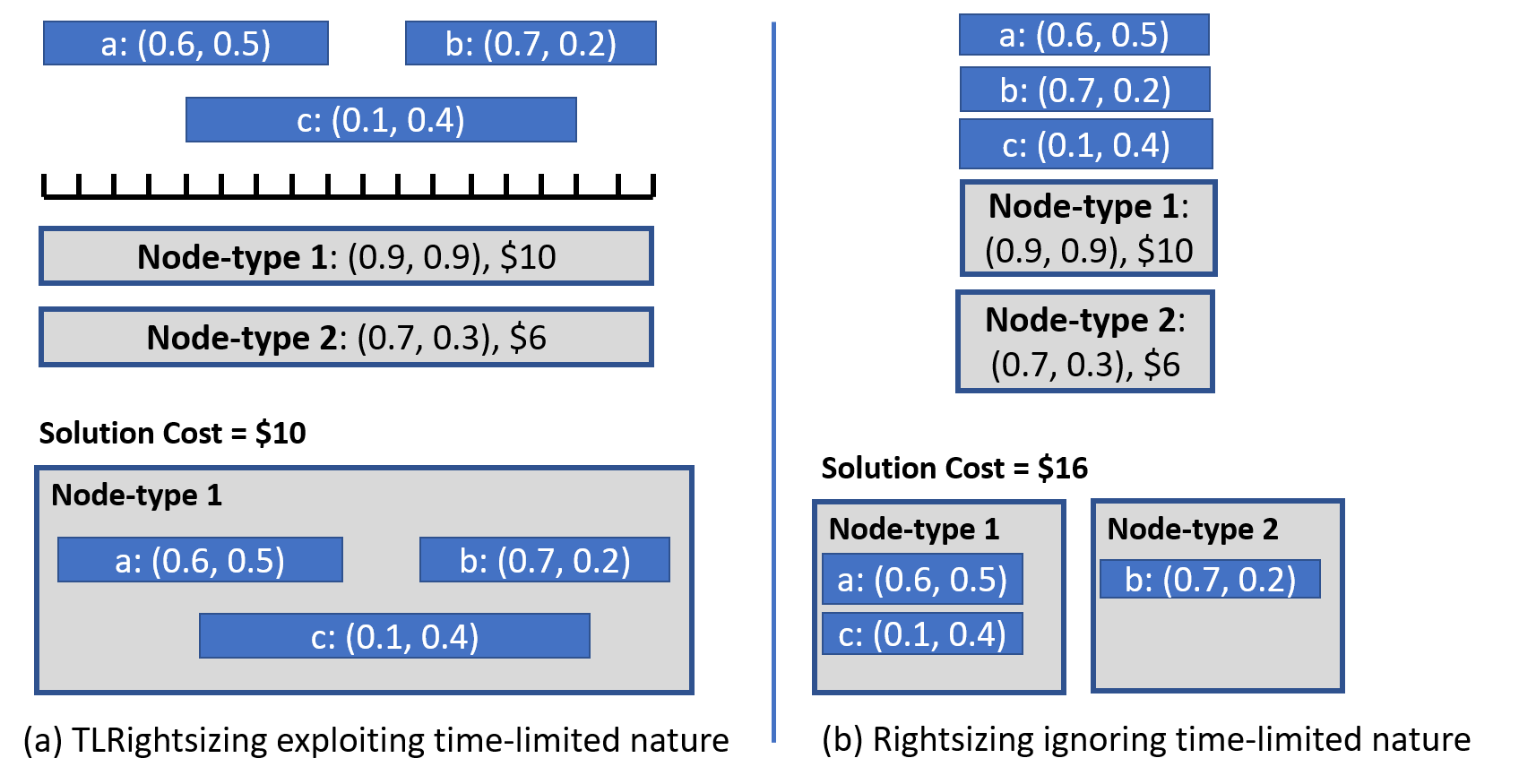}
\caption{Illustration of {\TLrightsize}. A simple example is shown with two resources ($D=2$), three tasks ($n=3$), and two node-types $(m=2)$.
The two dimensional demands and capacities, and node costs are shown.
Part (a) shows a  solution that exploits the time-limited nature and packs all the tasks in a single node of type 1 with cost $\$10$. 
In contrast, Part (b) ignores the
time-limited aspects, and shows the best solution that packs one node of each type with total cost $\$ 16$.
}
\label{fig:motivate}
\end{figure}

The problem, denoted {\TLrightsize}, starts with a set of tasks (with varying demands) and different nodes-types 
(with varying capacity and cost) as in {\Rightsize}. 
In addition, we assume a timeline divided into $T$ discrete timeslots, say a day divided into $T=24$ hours, 
and each task has start and end timeslots as an interval $[s, e]$. 
The objective of {\TLrightsize} is to purchase a minimum cost set of nodes and place the tasks on the nodes such 
that the aggregate demand of tasks placed on any node does not violate its capacity at any timeslot.
Figure \ref{fig:motivate} illustrates {\TLrightsize} and its benefits over {\Rightsize}.

\begin{figure}[tb]
    \centering
    \includegraphics[width=3.0in]{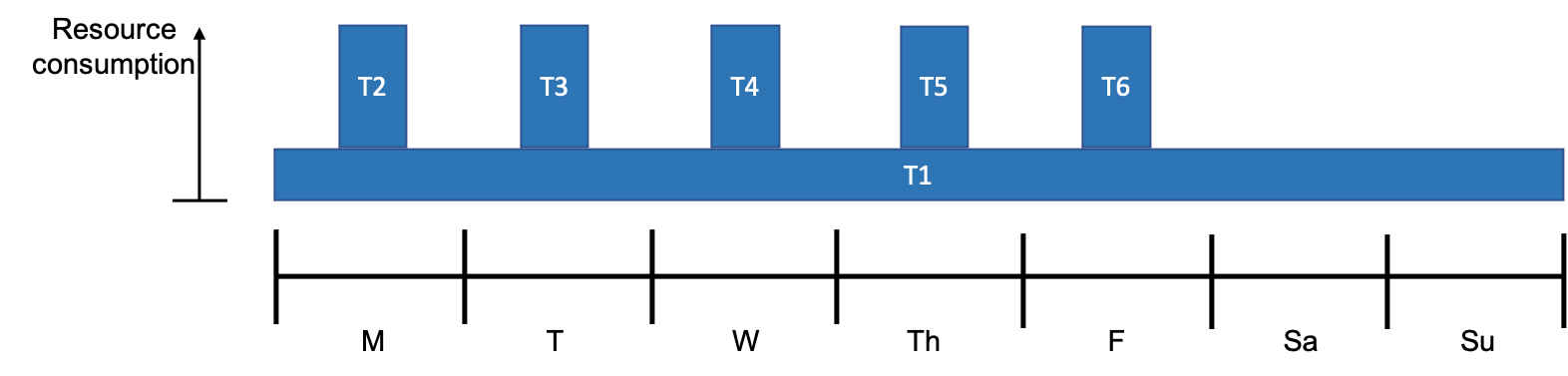}
    \caption{An original long-running task which serves stock market quotes may be modeled as six tasks, spanning a typical week: T1 models the low resource consumption regime, and T2-T6 model the additional demand during the 8-hour high resource consumption regimes during market open hours.}
    \label{fig:taskexample}
\end{figure}

For scenarios where demand, start and completion times are not directly known, we could build on cloud-centric 
time-series analysis ~\cite{khan2012workload, gmach2007workload} to leverage historical traces of task resource consumption from existing production or test deployments. The resource-demand of the task for various time windows (e.g. hourly, daily, weekly) can be estimated from the
traces and each window can be treated as an independent task with a specific start and ending time.
For example, a task handling stock market quotes may require high resource regime during market hours and low resource-demand regime at other times. For a timeline that spans a week starting Monday 00:00 hours, we can model this work as six tasks as shown in Figure~\ref{fig:taskexample}.

\mypara{Prior Work}
While {\TLrightsize} has not been studied in prior work, two streams of special cases,
{\Rightsize} and interval coloring, have been well explored.

{\it Rightsizing:} The {\Rightsize} problem corresponds to the restricted setting where all the tasks are perpetually active, in other words, 
there is no consideration of timeline ($T=1$). 

If tasks are items and nodes are bins, then the problem 
corresponds to multi-dimensional bin-packing over multiple bin-types. 
Since the problem is NP-hard, a polynomial time optimal solution is infeasible, unless NP=P. 
While exponential time integer programming~\cite{silva7} can produce optimal solutions, 
meta-heuristics such as genetic algorithms~\cite{silva13}, particle swarm \cite{silva14} and ant colony optimization \cite{silva15} 
can handle only small problem sizes due to high running time.

Alternatively, fast and scalable heuristics have been developed 
by building on strategies from the bin packing domain.
Working in the context of optimizing virtual machines, 
Panigrahy et al. \cite{panigrahy} study the single node-type special case ($m=1$), called vector bin packing.
They generalize bin packing heuristics (such as first and best-fit) and
demonstrate that the heuristics produce  near-optimal solutions in practice.
Gabay and Zaourar \cite{gabay} present similar procedures for multiple node-types with uniform costs.
Other approaches for special cases and variations of {\Rightsize} have been proposed (e.g., \cite{silva20, silva21})
and we refer to \cite{silva} for a detailed discussion.

We build on the work of  Silva et al. \cite{silva} and Patt-Shamir and Rawitz \cite{rawitz}, 
who present algorithms for the general {\Rightsize} problem.
The latter work proposes a two-phase procedure
that first maps each task to an appropriate node-type by comparing demands and capacities,
then solves for each node-type independently. They prove that the algorithm has an approximation ratio of $O(D)$.

From the theoretical perspective, Chekuri and Khanna \cite{chekuri-khanna} 
present an $O(\log D)$-approximation algorithm for the (single node-type) vector bin packing
and  Patt-Shamir and Rawitz \cite{rawitz} generalize the result to multiple node-types. 
However, the algorithms are based on an exponential size LP formulation and have a high running time of $n^{O(D)}$.

{\it Interval Coloring (or task scheduling):}
Interval coloring with bandwidths \cite{adamy} is equivalent to our setting with 
single dimensional resources ($D=1$) and single node-type ($m=1$).
In this problem, the input is a set of intervals with bandwidth requirements and the objective is to color them with 
minimum number of colors such that the aggregate bandwidth of intervals  of the same color 
does not exceed a fixed capacity at any timeslot. In our context, the intervals are tasks and the colors are nodes.
The problem is studied in both online and offline settings (see survey \cite{coloring-survey})
and procedures with $O(1)$ approximation ratios have been designed.

{\it Rightsizing for time-limited tasks:} The {\TLrightsize} problem generalizes and combines elements from the above two streams:
the notion of time-limited tasks from task scheduling, and the concepts of multi-dimensionality 
and multiple node-types from rightsizing.

\mypara{Our Contributions}
We introduce the {\TLrightsize} problem that reduces the cluster cost by exploiting the time-limited nature of tasks,
and make the following contributions:
\begin{itemize}
\item
We first design a two-phase algorithm, denoted {\penmap}, that serves as our baseline.
The algorithms uses ideas from  prior work on {\Rightsize} and task scheduling.
It first maps each task to an appropriate node-type via a penalty-based heuristic
and then solves each node-type separately. 
We present a detailed analysis and prove that the algorithm has an approximation ratio of $O(D\cdot \min(m, T))$.
\item
While {\penmap} algorithm performs reasonably well in practice, we identify certain deficiencies in the mapping strategy
and develop an improved mapping based on linear programming, denoted {\LPmap}.
\item
The {\LPmap} algorithm packs tasks mapped to each node-type in a maximal manner,
but we observe that it may leave empty capacity that can be filled by tasks mapped to other node-types.
We present a cross-node-type filling procedure that reduces the wastage and lowers the overall solution cost.
\item
We design a scalable strategy (that can handle large inputs) for determining a lowerbound on the optimal cost
and use it to measure the efficacy of our algorithms.
Using synthetic and real-world traces, we demonstrate that: 
(i) compared to the baseline {\penmap}, {\LPmap} offers up to $150\%$ reduction in the cost 
(normalized with respect to the lowerbound);
(ii) cross-node-type filling provides further improvement up to $8\%$, taking the solution closer to the optimal;
(iii) the final algorithm ({\LPmap} + cross node-type filling) generates solutions 
within at most $20\%$ of the lowerbound, and closer in most cases. 
\end{itemize}
\section{Problem Definition and Modeling}
\label{sec:defn}
\mypara{Problem Definition [{\TLrightsize}]}
Let the timeline be divided into $T$ discrete timeslots numbered $1, 2, \ldots, T$. Let $D$ be the number of {\em resources} (or dimensions). We have a set of $m$ {\em node-types} $\calB$. Each node-type $B\in \calB$ is associated with a price $\cost(B)$,
and offers a capacity $\ncap(B, d)$, along each dimension $d\in [1,D]$. The input consists of a 
set of $n$ tasks $\calU$; each task $u\in \calU$ is specified by a {\em demand} $\dem(u, d)$ along each dimension
$d$ and a {\em span} $[s(u), e(u)]\subseteq [1, T]$, with $s(u)$ and $e(u)$ being the {\em starting} and the {\em ending} timeslots of $u$.
We say that  task $u$ is {\em active} at a timeslot $t$, if $t\in [s(u), e(u)]$ and denote it as `$u\sim t$'.

A {\em $B$-type node} refers to a replica $b$ having the same price and capacity: $\cost(b)=\cost(B)$ and $\ncap(b, d) = \ncap(B, d)$, for all 
$d\in [1, D]$.
A feasible solution is to purchase a set of nodes $S$ and place each task in one of the nodes in $S$
such that for any node $b$, the total demand at any timeslot and any dimension does not violate the capacity offered by the node. 
\[
\forall (t\in [1,T], d\in [1, D]): \sum_{u\sim t~\&~u\in b} \dem(u, d)~\leq~\ncap(b,d),
\]
wherein $b$ is viewed as the set of tasks placed in it.
We refer to the above as the {\em capacity constraint}.
A solution may purchase multiple nodes of the same node-type.
The cost of the solution is the aggregate cost of the nodes purchased: $\cost(S) = \sum_{b\in S} \cost(b)$. The goal is to compute a solution of minimum cost.

\mypara{Timeline Trimming} 
While $T$ can be arbitrarily large, as is common in task scheduling (see e.g., \cite{bar-noy}), it can trimmed by considering 
only the start timeslots of tasks and ignoring the rest so that $T \leq n$. It is not difficult to argue that the 
transformation does not change the set of feasible solutions.

\section{Penalty-Based Mapping Algorithm}
\label{sec:penmap}
We design a baseline algorithm, denoted {\penmap}, by building on heuristics from prior work
on the following two special cases: interval coloring \cite{cloud-survey}, wherein number of node-types $m=1$
and dimensions $D=1$, and $\Rightsize$ \cite{silva, rawitz}, wherein there is no timeline ($T=1$).

\mypara{Two Phase Framework}
The intuition behind the algorithm is that each task must be placed on a node whose capacity matches the demand of the task.
For instance, placing a task $u$ having high demand along a particular dimension $d$ to a node with low capacity along $d$
would block us from accommodating other tasks on the node, leading to capacity wastage along the other dimensions.
The algorithm addresses the issue by using a two-phase framework. 
First,  each task $u\in \calU$ is mapped to a suitable node-type $B$
by comparing the demand of $u$ with the capacities of the node-types.
The second phase partitions the tasks based on their node-types and solves each node-type separately, as in the single node-type setting,
via a popular heuristic used in interval coloring.

\mypara{Mapping Phase} 
For a task $u$, define the relative demand (or height) with respect to a node-type $B$ as:
\begin{eqnarray*}
\havg(u|B) = \frac{1}{D}\sum_{d\in [1,D]} \frac{\dem(u, d)}{\ncap(B, d)}
\end{eqnarray*}
Intuitively, $\havg(u|B)$ is a measure on the space occupied by $u$, if the task were placed in a node of type $B$. 
Taking the cost into consideration, define the {\em penalty of $u$ relative to $B$} as: 
\[
\pavg(u|B) = \cost(B)\cdot \havg(u|B)
\]
We map $u$ to the node-type $\bavgstr(u)$, yielding the least penalty:
\[
\pavgstr(u) = \min_{B\in \calB}~\pavg(u|B) \mbox{ \& } \bavgstr(u) = \argmin_{B\in \calB}~\pavg(u|B)
\]

\mypara{Placement Phase}
We partition the tasks based on their node-types and process each group separately.  
Consider a node-type $B$ and let $V_B$ denote the set of tasks mapped to $B$.
We process the tasks in $V_B$ in the {\em increasing order of starting timeslots}.
Suppose we have processed a certain number of tasks and have purchased a set of nodes to accommodate them.
Let $u$ be the next task. We check if $u$ can fit in some node $b$ purchased already without violating the capacity of the node. 
If so, among the feasible nodes, we place $u$ in the node purchased the earliest (first-fit).
Otherwise, we purchase a new $B$-type node and place $u$ in it.
Figure \ref{fig:penmap} shows a pseudocode. 

\mypara{Alternative Mapping and Fitting Policies}
An alternative mapping policy (see \cite{rawitz}) is to define the relative demand as the maximum over 
the dimensions (as against average), and define the penalty and best node-type analogously:
\[
\hmax(u|B) = \max_{d\in [1,D]} \frac{\dem(u, d)}{\ncap(B, d)}.
\]

An alternative to the first-fit policy is the more refined {\em similarity-fit}
that emulates the best-fit strategy (adapted from a dot-product strategy \cite{panigrahy, gabay}). It tries to minimize the capacity wastage 
by selecting the node $b$ (among the feasible nodes) whose remaining capacity is most similar (dot-product shown below) to the demand of $u$.
\[
\sum_{t\in [s(u), e(u)]} \sum_d \frac{\dem(u,d)}{\ncap(B, d)} \times \frac{\rem(b, d|t)}{\ncap(B, d)},
\]
where $\rem(b, d|t)$ is the remaining capacity of $b$ along dimension $d$ at timeslot $t$. 
We can derive cosine similarity via dividing the product by the norms of capacity-normalized demand and 
remaining capacity vectors involved in the above inner product. Then, the strategy is to place $u$ in the node 
offering the maximum cosine similarity value.

\mypara{Time Complexity}
Time taken for computing the mapping is $O(n\cdot m)$. Time required to test whether a task fits a node is $O(D \cdot T)$.
Given that the number of nodes can be at most $|S|$, the total number of nodes purchased, overall running time of the algorithm is
$O(n\cdot m + n\cdot |S|\cdot D \cdot T)$.
Typically the last term of this bound dominates in practice.

\begin{figure}[t]
\begin{center}
\begin{boxedminipage}{\hsize}
\begin{tabbing}
xx\=xx\=xx\=xx\=xx\=xx\=xx\=\kill
{\bf Mapping Phase:}\\
For each task $u$ and each node-type $B$\\ 
\> $\havg(u|B) = \frac{1}{D}\sum_d \frac{\dem(u, d)}{\ncap(B, d)}$\\
\> $\pavg(u|B) = \cost(B)\cdot \havg(u|B)$.\\
For each task $u$:\\
\> $\pavgstr(u) = \min_B~\pavg(u|B)$.\\
\> Map $u$ to $\bavgstr(u) = \argmin_B~\pavg(u|B)$\\
For each node-type $B$:\\
\> Let $\VBpen$ be the set of tasks mapped to $B$.\\
{\bf Placement Phase:}\\
For each node-type $B$:\\
\> Initialize solution $S_B=\emptyset$\\
\> Sort $\VBpen$ in the increasing order of starting timeslots.\\
\> For each task $u$ in the above order:\\
\> \> If $u$ can fit in some node in $S_B$:\\
\> \> \> Among the feasible nodes, \\
\> \> \> \> place $u$ in the node purchased the earliest.\\
\> \> Else:\\
\> \> \> Purchase a new $B$-type node $b$ and place $u$ in it.\\
Output: $\Spen = \bigcup_B S_B$.
\end{tabbing}
\end{boxedminipage}
\end{center}
\caption{Algorithm \penmap}
\label{fig:penmap}
\end{figure}

\section{{\penmap} : Analysis}
For the special case of interval coloring ($m=1$, $D=1$), 
prior work \cite{round-ufp} derives an $O(1)$ approximation ratio.
Similarly, for the special case of $\Rightsize$ ($T=1$),
an $O(D)$ approximation ratio is implicit in existing literature \cite{rawitz}.
We next present an analysis for the general {\TLrightsize} setting
and prove that the {\penmap} algorithm has an approximation ratio of
$O(D\cdot \min(m, T))$. The result matches and generalizes both the prior special cases.

For the sake of concreteness, we focus on the $\havg$-based mapping and the first-fit strategy, 
but the analysis can easily be adapted to the combinations involving $\hmax$ and similarity-fit.
We say that a task $u$ is {\em small}, if for all node-types $B$ and all dimensions $d$, $\dem(u, d) \leq \ncap(u, B)/2$.
We  first work under the practically reasonable assumption that  all the tasks are small and then discuss the general case.

Let $\Sopt$ and $\Spen$ denote the optimal and {\penmap} solutions, respectively.
We derive a lower-bound on $\cost(\Sopt)$ and an upper-bound on $\cost(\Spen)$, based on the notion of congestion, defined next.
Given a subset of tasks $U\subseteq \calU$, define the {\em congestion induced by $U$} as
the maxima over the timeline of the aggregate minimum penalties of tasks active at each timeslot:
\[
\cong(U) = \max_t \sum_{u\in U, u\sim t}\pavgstr(u).
\]
The following lemma shows that $\cost(\opt)$ is at least the congestion induced by the entire task set $\calU$.

\begin{lemma}
\label{lem:aaa}
$\cost(\opt) \geq \cong(\calU)$.
\end{lemma}
\proof
Consider any node-type $B$. Let $N_B$ denote the number of $B$-type nodes purchased by $\opt$,
and let $\VBopt$ denote the set of tasks placed in $B$-type nodes. 
For any timeslot $t$ and dimension $d$, the aggregate demand is given by:
\[
\sum_{u\in \VBopt, u\sim t} \dem(u, d)
\]
To accommodate this demand, the number of $B$-type nodes $N_B$ must be at least:
\[
\left \lceil \frac{\sum_{u\in \VBopt, u\sim t} \dem(u, d)}{\ncap(B, d)}\right \rceil.
\]
Taking maximum over all $(t, d)$ pairs, we get that $N_B$ must be at least:
\[
\max_{t,d} \sum_{u\in \VBopt, u\sim t} \frac{\dem(u, d)}{\ncap(B, d)}
\geq
\max_t \sum_{u\in \VBopt, u\sim t} \havg(u|B),
\]
where we replace the maxima over $d$ by the average. Hence,
\begin{eqnarray*}
\cost(\Sopt) 
&= &\sum_B \cost(B)\cdot N_B\\
&\geq &\sum_B \cost(B)\max_t \sum_{u\in \VBopt, u\sim t} \havg(u|B)\\
&\geq& \sum_B \max_t \sum_{u\in \VBopt, u\sim t} \pavg(u|B)\\
&\geq& \sum_B \max_t \sum_{u\in \VBopt, u\sim t} \pavgstr(u)\\
&\geq& \max_t \sum_B \sum_{u\in \VBopt, u\sim t} \pavgstr(u)\\
&\geq& \max_t \sum_{u\sim t} \pavgstr(u),
\end{eqnarray*}
where the last inequality follows from the fact that every task active at $t$ is placed in a node of some type.
\qed


We next analyze the solution $\Spen$.
Let $\cost(\calB)$ denote the sum of costs of the node-types: $\cost(\calB)=\sum_B \cost(B)$.
For a node-type $B$, let $\VBpen$ denote the set of tasks mapped to $B$ by the solution.
In contrast to the lower-bound on $\cost(\opt)$ which is in terms of the congestion over the entire task set,
we prove a (comparatively weaker) upper-bound on $\cost(\Spen)$ in terms 
of the summation of the congestion induced within each node-type.

\begin{lemma} 
\label{lem:bbb}
Assuming all the tasks are small, 
\[
\cost(\Spen) \leq \cost(\calB) + (2D)\cdot \Sigma_B \cong(\VBpen).
\]
\end{lemma}
\proof
Consider any node-type $B$ and let $N_B$ denote the number of $B$-type nodes purchased by $\Spen$.
We claim that:
\begin{eqnarray}
\label{eqn:eee}
|N_B| \leq 1 + (2D)\cdot \max_t \sum_{u\in \VBpen, u\sim t} \havg(u|B).
\end{eqnarray}
The proof is by induction. Consider the processing of a task $v$ and let $S$ denote the set of nodes purchased already.
Assume by induction that $|S|$ satisfies the bound before processing $v$ and we shall show that it remains true
after the processing. If $v$ could fit in one of the nodes in $S$, then no new node is purchased and the bound
remains true. 

Suppose $v$ could not fit into any of the nodes in $S$ and a new node is purchased. 
This means that for any 
node $b\in S$, placing $v$ in $b$ would result in violation of the capacity along some dimension $\hatd$ at some timeslot $\hatt$
within the span of $v$:
\[
\dem(v, \hatd) + \sum_{u\in b, u\sim \hatt} \dem(u, \hatd) \geq \ncap(b, \hatd).
\]

\eat{
\begin{figure}
\centering
\includegraphics[width=2.5in]{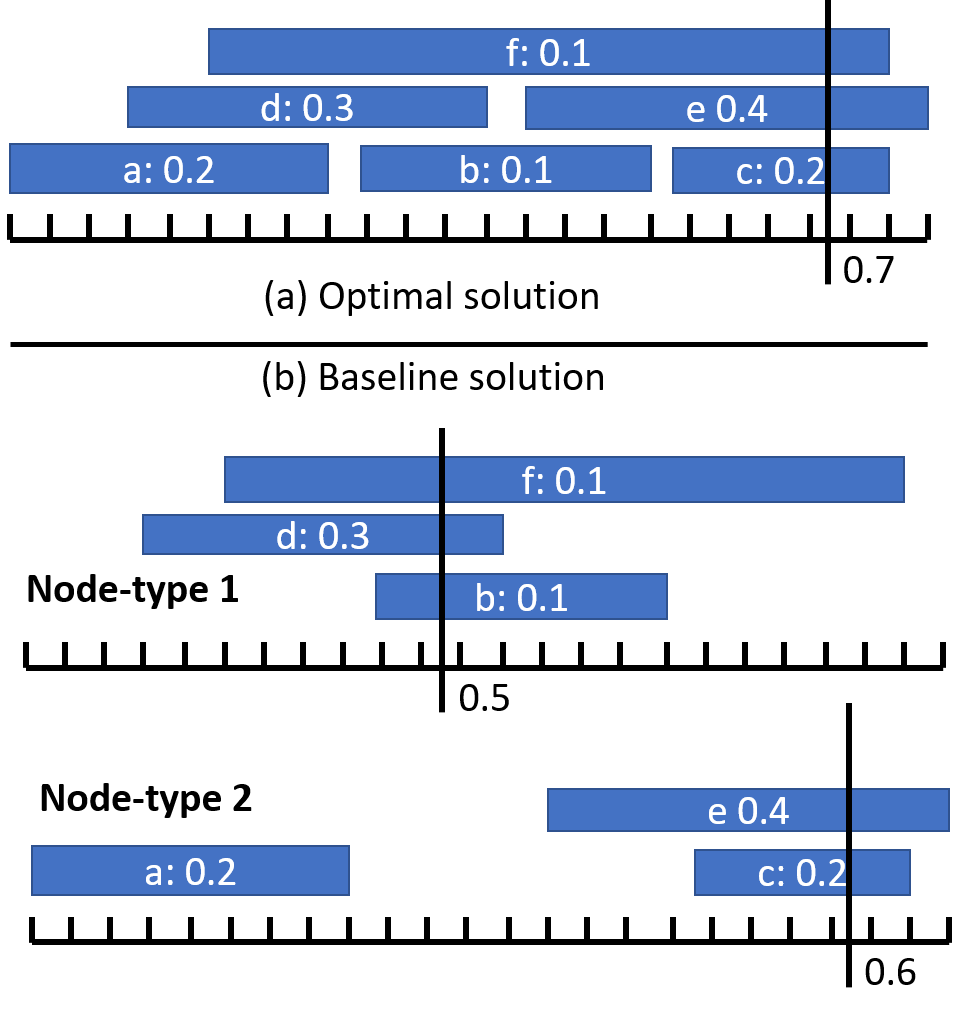}
\caption{
Illustration of congestion. Penalties $\pavgstr(u)$ are shown. 
The lower-bound on $\cost(\Sopt)$ is the maximum aggregate penalty over the timeline; in this example, $0.7$.
The upper-bound on $\cost(\Spen)$ is sum of the maxima over the node-types. This illustration assumes 
that there are two node-types and that the algorithm maps $\{b,d,f\}$ to the first and
$\{a,c,e\}$ to the second, making the sum $0.5+0.6=1.1$.
}
\label{fig:congestion}
\end{figure}
}

Since the tasks are processed in the increasing order of starting timeslots, all the above tasks $u$ 
must start earlier than $v$ and so, they all must be active at the starting timeslot $s(v)$, because they overlap with $v$. 
Consequently, we can assume without loss of generality that $\hatt = s(v)$. 
Thus,
\[
\dem(v, \hatd) + \sum_{u\in b, u\sim s(v)} \dem(u, \hatd) \geq \ncap(b, \hatd).
\]
Since we assume that the tasks are small, $\dem(v, \hatd) \leq \ncap(b, \hatd)/2$ and hence:
\[
\sum_{u\in b, u\sim s(v)} \dem(u, \hatd) \geq \ncap(b, \hatd)/2,
\]
or equivalently:
\begin{eqnarray*}
1/2 
&\leq& \sum_{u\in b, u\sim s(v)} \frac{\dem(u, \hatd)}{\ncap(b,\hatd)} \\
&\leq& \sum_{u\in b, u\sim s(v)} \sum_d \frac{\dem(u, d)}{\ncap(b,d)} \\
&=& D\cdot \sum_{u\in b, u\sim s(v)} \frac{1}{D}\sum_d \frac{\dem(u, \hatd)}{\ncap(b,d)} \\
&=& D\cdot \sum_{u\in b, u\sim s(v)} \havg(u|B).
\end{eqnarray*}

Summing over all the nodes in $S$:
\[
\frac{|S|}{2}
\leq
D \cdot \sum_{b\in S} \sum_{u\in b, u\sim s(v)} \havg(u|B).
\]
Taking $U'$ as set of tasks included in $S$ (i.e., the set of tasks processed before $v$),
we get:
\[
|S| 
\leq
(2D)\cdot \sum_{u\in U', u\sim s(v)} \havg(u|B).
\]
The above inequality considers the specific timeslot $s(v)$. 
We complete the induction step by relaxing it to the timeslot yielding the maximum summation, observing that $U' \subseteq \VBpen$,
and accounting for the extra node purchased for placing $v$.
The claim given in (\ref{eqn:eee}) is proved.

We are now ready to prove the lemma. Cost of $\Spen$ is:
\begin{eqnarray*}
&=& \sum_B \cost(B) \cdot N_B\\
&\leq& \cost(\calB) + (2D)\cdot \sum_B \max_t \sum_{u\in \VBpen, u\sim t} \pavg(u|B)\\
&=& \cost(\calB) + (2D)\cdot \sum_B \max_t \sum_{u\in \VBpen, u\sim t} \pavgstr(u),
\end{eqnarray*}
where the second inequality is true because
the algorithm maps each task $u$ to the node-type offering the least penalty.
\qed

Combing the upper and the lower-bounds, we next prove an approximation ratio for the {\penmap} procedure.

\begin{theorem}
Assuming that all the tasks are small,
\[
\cost(\Spen) \leq \cost(\calB) + (2D)\cdot \min(m, T)\cdot \cost(\Sopt)
\]
\end{theorem}
\proof
Referring to Lemma \ref{lem:bbb}, we have that:
\begin{eqnarray}
\nonumber
\sum_B \cong(\VBpen) 
&\leq& \sum_B \cong(\calU) \\
\label{eqn:fff}
&=& m\cdot \cong(\calU) \leq m\cdot \cost(\opt), \ \ \ \ \ \ \ 
\end{eqnarray}
where the first inequality is true because $\VBpen \subseteq \calU$
and the last inequality follows from Lemma \ref{lem:aaa}.
Furthermore, 
\[
\sum_B \cong(\VBpen) 
=
\sum_B \max_t \sum_{u\in \VBpen, u\sim t} \pavgstr(u) \\
\leq
\sum_{u\in \calU} \pavgstr(u),
\]
where the inequality is true because every task contributes at most once in the summation.
On the other hand,
\begin{eqnarray*}
\cost(\Sopt) 
&\geq&
\max_t \sum_{u\in \calU, u\sim t} \pavgstr(u) \\
&\geq& \frac{1}{T}\sum_t \sum_{u\in \calU, u\sim t} \pavgstr(u) 
\geq \frac{1}{T}\sum_{u\in \calU} \pavgstr(u),
\end{eqnarray*}
where first inequality follows from Lemma \ref{lem:aaa} and the last inequality from the fact that every task is active at some timeslot.
Hence,
\begin{eqnarray}
\label{eqn:ggg}
\sum_B \cong(\VBpen) 
\leq
\sum_{u\in \calU} \pavgstr(u) 
\leq
T\cdot \cost(\opt).
\end{eqnarray}
The theorem is proved by taking minimum of the bounds given by (\ref{eqn:fff}) and (\ref{eqn:ggg}),
and appealing to Lemma \ref{lem:bbb}.
\qed

The theorem provides an approximation ratio of $O(D\cdot \min(T, m))$, for the case of small tasks.
To handle the general case, we segregate the large tasks and apply {\penmap} separately
and derive the final solution by taking the union of the two solutions. 
Using an analysis similar to that of small tasks, we can show that the case of large tasks also 
admits an approximation ratio of $O(D\cdot \min(T, m))$, thereby yielding the same overall ratio for the general case
(but with an increased hidden constant).
However, our experiments show that in practice,  the solutions are much closer to the optimal,
even without segregating the tasks into the two classes.

\begin{figure}
\centering
\includegraphics[width=3.0in]{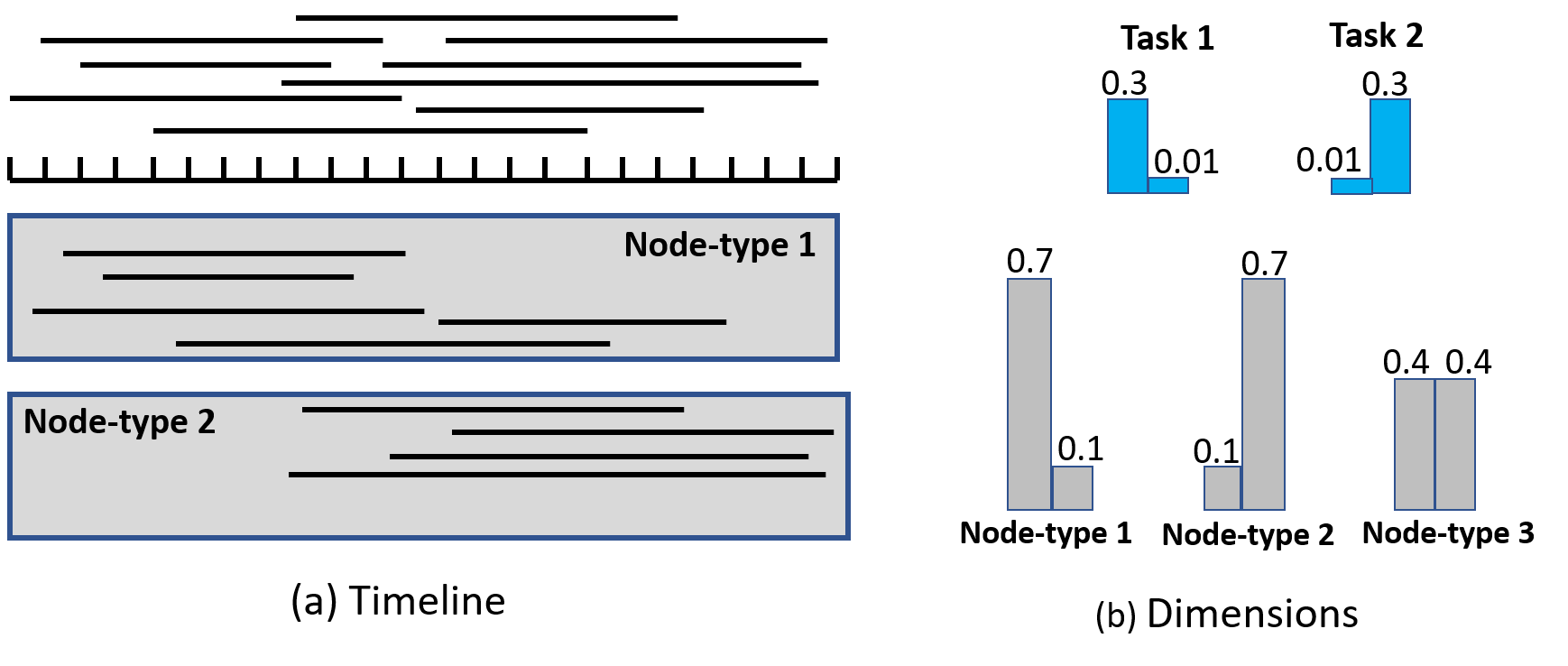}
\caption{Illustration of deficiencies of penalty-based mapping.
In part (a), tasks  mapped to the first node-type are mainly 
active in the earlier segments of the timeline, whereas those mapped to the second are mainly active in the later segments, 
leading to resource wastage.
In part (b), taking $D=2, m=3$ and $T=1$, resource demands/capacities along the two dimensions are shown.
{\penmap} would map tasks 1 and 2 to node-types 1 and 2, respectively, 
thereby utilizing two nodes, one from each type,
whereas both can be accommodated in a single node of node-type 3.
}
\label{fig:deficiency}
\end{figure}

\section{Linear Programming Based Mapping}
While the penalty based mapping works reasonably well in practice, 
there are certain deficiencies in the strategy and we design an improved mapping based on linear programming.

\subsection{Penalty Based Mapping: Deficiencies}
The penalty-based strategy makes the node-type choices for the tasks independent of each other,
optimizing on an individual basis without considering the interactions. 
The lack of a collective analysis may lead to wastage of resource capacities,
resulting in sub-optimal packing and higher cost. 

There are two aspects to the issue, pertaining to the timeline and the resource dimensions.
Firstly, tasks mapped to a node-type may be predominantly active
only at certain segments on the timeline and less active at others. 
When the group of tasks get placed in nodes of the given type,
the resource capacity gets wasted at the latter segments.
Secondly, tasks with high demand along a particular dimension $d$ (say memory)
are likely to get mapped to a node-type with higher capacity along $d$,
wasting the capacity along the other dimensions. 
Figure \ref{fig:deficiency} illustrates the two aspects. 
In addition, the node-type is selected based on normalized demands in an aggregated manner,
ignoring the effect of individual dimensions.

\eat{
\begin{figure}[t]
\begin{center}
\begin{boxedminipage}{\hsize}
\begin{tabbing}
xx\=xx\=xx\=xx\=xx\=xx\=xx\=\kill
{\bf Mapping Phase:}\\
Construct the linear program and derive the solution $x^*$.\\
For each task $u$:\\
\> Map $u$ to $\pi_{\LP}(u) = \argmax_B~x^*(u, B)$\\
For each node-type $B$:\\
\> Let $\VBlp$ be the set of tasks mapped to $B$.\\
{\bf Placement Phase:}\\
\> Same as {\penmap} (Figure \ref{fig:penmap}), but\\
\> \> use the partitioning $\VBlp$ instead of $\VBpen$.\\
Output Solution: $\Slp$.
\end{tabbing}
\end{boxedminipage}
\end{center}
\caption{Algorithm \LPmap}
\label{fig:LPmap}
\end{figure}
}

\subsection{Linear Programming}
\label{sec:lp-lb}
We can address the above issues and find an improved mapping via linear programming (LP).
Let $\pi:\calU\rightarrow \calB$ be any task to node-type mapping. 
Consider executing the two-phase algorithm (Figure \ref{fig:penmap})
with the given mapping $\pi$ (instead of penalty-based mapping),
followed by the same greedy placement for each node-type.
Let $\Spi$ be the solution output by the algorithm. 
Using a proof similar to that Lemma \ref{lem:aaa},
we can show the following lower-bound, denoted $\LB(\pi)$:
\[
\cost(\Spi) \geq \sum_B \max_{t,d} \sum_{u\in \VBpi, u\sim t} \frac{\dem(u, d)}{\ncap(B, d)},
\]
where $\VBpi$ is the set of tasks mapped to $B$ under $\pi$. 
Then, $\cost(\opt)$ is at least the minimum lower-bound over all the possible mappings:
\[
\cost(\opt) \geq \min_{\pi} \LB(\pi).
\]

Our goal is to find the mapping $\pi$ with the least lower-bound $\LB(\pi)$. 
The goal can be accomplished via  integer programming.
We introduce a variable $x(u, B)$ for each pair of task and node-type, 
representing whether $u$ is mapped to $B$. 
We add a variable $\alpha_B$ for each node-type $B$,
capturing the contribution from the node type to the lower-bound:
\[
\sum_B \max_{t,d} \sum_{u\in \VBpi, u\sim t} \frac{\dem(u, d)}{\ncap(B, d)}.
\]

The integer program is shown below.
\begin{eqnarray}
\nonumber
\min && \sum_B \cost(B)\cdot \alpha_B\\
\label{eqn:aaa}
\forall u: && \sum_B x(u, B) = 1\\
\label{eqn:bbb}
\forall (B, t,d): && \sum_{u\sim t} x(u, B) \cdot \frac{\dem(u, d)}{\ncap(B,d)} \leq \alpha_B \\
\label{eqn:ccc}
\forall (u,B): && x(u, B) \in \{0,1\},
\end{eqnarray}
where the first and the last constraint ensure that $u$ is mapped to exactly one node-type.
Once the values of $x(u, B)$ are fixed,
the optimal value for $\alpha_B$ is automatically the contribution of $B$.
Thus, the integer program yields the desired mapping.

Unfortunately, it is prohibitively expensive to solve integer programs even on moderate size inputs. 
So, we resort to linear programming by relaxing the integrality constraint (\ref{eqn:ccc}) as:
\begin{eqnarray}
\label{eqn:ddd}
\forall (u,B): \quad 0 \leq x(u, B) \leq 1.
\end{eqnarray}
Let $x^*(u, B)$ be the optimal LP solution. The objective value provides a lower-bound on $\cost(\opt)$.

\subsection{LP Rounding}
In contrast to the integer program, LP produces a fractional solution:
for each task $u$, the solution consists of a distribution (vector) over the node-types 
$[x^*(u, B)]$ summing to $1$, where $x^*(u, B)$ is the fractional extent to which $u$ is assigned to $B$.
We design a procedure for transforming (rounding) the fractional solution into an integral solution.
The procedure is based on a critical observation that the solution $x^*$ would be nearly integral,
when $n$ is large enough.

\begin{lemma}
In the solution $x^*$, the number of variables $x^*(u, B)$ with non-integral values is at most $n+mTD$.
\end{lemma}
\proof
Assume that $x^*$ is an extreme point (or basic feasible) solution.
The number of variables is $nm + m$ and so, at least as many constraints have to be tight \cite{lpbook}.
Of these, $n + mTD$ can be from the constraints (\ref{eqn:aaa}) and (\ref{eqn:bbb}),
and the rest must come from (\ref{eqn:ddd}). Hence, at most $(n + mTD - m)$ of the $nm$ variables of the form $x(u, B)$ can be proper fractions. 
\qed

\begin{figure}
\centering
\includegraphics[width=2.5in]{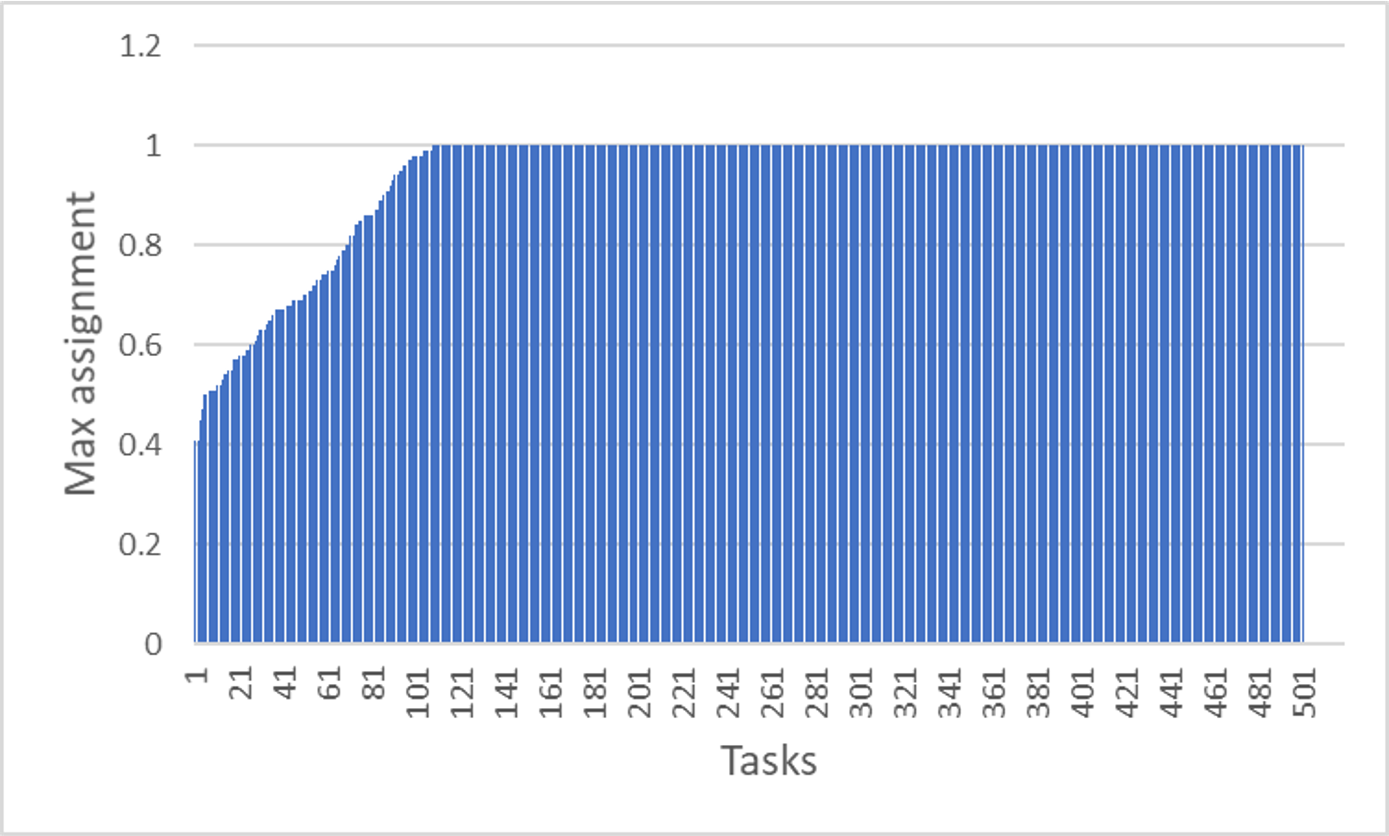}
\caption{
Illustration for near-integrality using a sample input from our experimental study with $n=500, m=10, D=5$ and $T=24$.
The X-axis shows the tasks $u$ and the Y-axis shows $x_{\max}(u)$. Tasks are sorted by their value for the ease of visualization.
We see that most of the tasks are assigned to a single node-type or show a clear preference for a particular node-type (the one receiving
the maximum assignment)
}
\label{fig:integral}
\end{figure}

The lemma shows that of the $n\cdot m$ variables $x^*(u, B)$, most would be integral, when $n\gg TD$.
From our experimental study, we observe that the above phenomenon manifests strongly in practice 
(irrespective of the values of $n$ and $TD$).
For a task $u$, let $x_{\max}(u) = \max_B x^*(u)$ denote the maximum extent to which the task is assigned to a node-type.
In the worst case, the solution can assign $u$ to an extent of $1/m$ to each node-type.
However, the results show that $x_{\max}(u)=1$ or close to $1$ for vast majority of tasks.
Meaning most of the tasks get assigned to a single node-type, or show a clear preference for a particular node-type.
Figure \ref{fig:integral} provides an illustration by considering a sample input from the experimental study.

Based on the above observation, we employ a simple, but effective, rounding heuristic:
map each task $u$ to the node-type $B$ to which the task is assigned to the maximum extent:
\[
\pi_{\rm LP}(u) = \argmax_B \quad x^*(u, B).
\]
Given the mapping, we partition the tasks according to the node-types and 
treat each node-type independently,
using the same greedy placement heuristic as in {\penmap} and obtain a solution $\Slp$.

\mypara{Analysis}
The analysis of {\LPmap} procedure is similar to that of {\penmap} and a sketch is provided below.
For any task $u$, there must be at least one node-type $B$ with $x^*(u, B) \geq 1/m$.
Thus, the rounding incurs a loss by a factor of at most $m$. This fact combined with
a proof similar to that of {\penmap} can be used to show that
that solution output by the {\LPmap} satisfies:
\[
\cost(\Slp) \leq \cost(\calB) + (2Dm)\cdot \cost(\opt),
\]
yielding an approximation ratio of $O(Dm)$. The solutions are much closer to the 
optimal in practice, with the margin being no more than $30\%$ in our experiments.

\begin{figure}[t]
\begin{center}
\begin{boxedminipage}{\hsize}
\begin{tabbing}
xx\=xx\=xx\=xx\=xx\=xx\=xx\=\kill
{\bf Mapping Phase:}\\
Construct the linear program and derive the solution $x^*$.\\
For each task $u$:\\
\> Map $u$ to $\pi_{\LP}(u) = \argmax_B~x^*(u, B)$\\
{\bf Placement Phase:}\\
$R = \calU$ \quad\quad\quad // Remaining tasks.\\
Sort the node-types $B$ in decreasing order of $\frac{\sum_d \ncap(B, d)}{\cost(B)}$. \\
For each node-type $B$ in the above order:\\
\> Initialize solution $S_B=\emptyset$\\
\> Let $\VBlp$ be the set of tasks mapped to $B$.\\
\> Let $U = \VBlp\cap R$\quad\quad // Mapped to $B$ and remaining\\
\> Sort $U$ in the increasing order of start timeslots.\\
\> For each task $u$ in the above order:\\
\> \> If $u$ can fit in some node in $S_B$:\\
\> \> \> Among the feasible nodes, \\
\> \> \> \> place $u$ in the node purchased the earliest.\\
\> \> Else:\\
\> \> \> Purchase a new $B$-type node $b$ and place $u$ in it.\\
\> // Cross node-type filling or piggy-backing\\
\> Sort the tasks $u$ in $R$ in increasing order of $\havg(u, B)$. \\
\> For each task $u$ in $R$ in the above order\\
\> \> If $u$ can fit in some node in $S_B$: Among them\\
\> \> \> place $u$ in the node purchased the earliest.\\
\> Delete all the tasks placed in this iteration from $R$\\
Output: $\Slp = \bigcup_B S_B$.
\end{tabbing}
\end{boxedminipage}
\end{center}
\caption{Algorithm {\LPmap} with cross node-type filling}
\label{fig:piggy}
\end{figure}

\subsection{Cross Node-Type Filling}
\label{sec:cross}
The placement procedure is maximal in the sense that for any node-type $B$, it opens a new node for a task $u$
only when it cannot fit any of the nodes purchased already. However, the heuristic may leave empty spaces
that can be filled by tasks of {\em other} node-types.
We present a cross node-type filling method that reduces the overall cost 
by placing tasks mapped to the other node-types in the above empty spaces.

Suppose we have processed node-types $B_1, B_2, \ldots, B_k$ and let $S$ denote the set of nodes purchased already. 
Let $R$ denote the set of remaining tasks that are yet to be placed (these would be mapped to node-types $\geq k+1$).
We wish to place as many tasks from $R$ in the nodes found in $S$. We employ a greedy heuristic for this purpose.
We sort the tasks $u\in R$ in the increasing order of $\havg(u, B_k)$, intuitively, 
arranging the task in terms of the space they would occupy.
For each task $u$ in the above order, we check if $u$
can be placed in some node in $S$ without violating the feasibility constraint and if so, we place it in node purchased the earliest.
The process is repeated till all the tasks are placed.

In the above process, tasks mapped to a node-type $B$ get an opportunity to piggy-back
on the nodes purchased for the earlier node-types. 
Thus, the tasks mapped to later node-types are more likely to piggy-back 
and so, the  ordering of the node-types is important. 
We sort the node-types in the decreasing order of capacity to cost ratio:
$
\sum_d \ncap(B, d)/\cost(B).
$
Intuitively, the ratio measures the capacity offered per unit cost 
and the sorting puts less cost-effective node-types later in the ordering, giving more opportunity for their tasks to piggy-back.
A pseudo-code is presented in Figure \ref{fig:piggy}. For each node-type $B$,
the tasks that are mapped to $B$ are first placed, 
and then, the remaining tasks (mapped to later node-types) are accommodated as much as possible.
The cross node-type filling method can also be applied to penalty based mapping (or to any other mapping strategy).

\mypara{Time Complexity}
The LP is designed to only find the task to node-type mapping, rather than a full solution
encoding the placement of tasks in nodes. 
Consequently, the size of the LP is fairly small with the number of variables constraints being $O(nm)$
and $O(mTD)$, respectively. Interior point methods can converge in polynomial time
and the execution time  depends on the solver's performance.
Given the LP solution, the time complexity of the rest of the procedure is similar to that of {\penmap}:
the mapping takes time $O(n\cdot m)$, whereas placement takes time $O(n\cdot |S|\cdot D\cdot T)$,
resulting in a total execution time of $O(n\cdot m + n\cdot |S|\cdot D \cdot T)$.

\section{Experiments}
\label{sec:expt}


\subsection{Set-up}
\mypara{System}
A machine with 2.5 GHz, Intel Core i7, 16GB memory and running macOS version 10.15.2 was used for running 
experiments implemented in python 3.8.7. We use python-mip package \cite{pymip} with CBC solver for solving LP.

\mypara{Google Cloud Trace 2019 (GCT 2019)} 
{\google} is used for real-world evaluation. 
10M collection events (begin and end events of groups of tasks) and all machine-types (node-types) available for a single cluster (labeled ``a'')~\cite{gct2019} were sampled using BigQuery. 
Demand and capacity are two-dimensional (CPU and memory) and normalized in the trace.
Entries with missing fields are purged from sampled trace. Time-stamps are converted to seconds. 
Task start and end times are discovered using task creation and end events. 
Thus, the task intervals, demands and capacities are drawn from the real trace.
The node-type cost is generated synthetically (see below), and  using publicly available pricing coefficients~\cite{gct2019pm}.
The processed data contains about 13K tasks and 13 node-types. Given $n$ and $m$, which is one experimental scenario, we construct an input instance
by randomly sampling $n$ tasks and $m$ node-types from this processed data. Results are average across $5$ such random input instances per scenario.
Timeline trimming is performed as discussed in Section \ref{sec:defn}.

\begin{table}[t]
\caption{Default values for parameters}
\begin{center}
\begin{tabular}{|c||c|c|}
\hline
\textbf{Parameter} & \textbf{Traces} & \textbf{\textit{Value}} \\
\hline
$n$ & Both & 1000 \\
\hline
$m$  & Both & 10 \\
\hline
T & Synthetic & 24 \\
\hline
Capacity & Synthetic & [0.2, 1.0] \\
\hline
Demand & Synthetic & [0.01, 0.1] \\
\hline
Dimensions & Synthetic & 5 \\
\hline
\end{tabular}
\label{tbl:defaultparams}
\end{center}
\vspace{-0.2in}
\end{table}

\mypara{Synthetic Data}
The {\google} trace is $2$-dimensional and the task demands are fixed and small compared to node-capacities.
In order to explore higher dimensions and different demand categories,
we setup a synthetic benchmark using a random generator with inputs $n$, $m$, $D$, $T$, 
and intervals for demand and capacity of the form $[a,b] \subseteq [0,1]$.
Each of the $D$ components of demand and capacity is uniformly and independently selected from its respective interval.
For each task $u$, $[s(u), e(u)]$ is uniformly selected from $[1, T]$. 
We fixed $T=24$ and capacity interval as $[0.2, 1]$. We pick $D\in \{2,5,7\}$, $m\in \{5,10,15\}$,
demand intervals from $\{[0.01, 0.05], [0.01, 0.1], [0.01, 0.2]\}$, and $n$ as per the experiment.
Unless specified, default values used in specific experiments are shown in Table \ref{tbl:defaultparams}.
As in {\google}, for each scenario, results are averaged over 5 random inputs
and timeline trimming is performed.

\mypara{Cost-model}
Node-type cost is computed as follows: 
\begin{equation}
\label{eq:cm}
cost(B) = \sum_{d \in [1,D]}{c_d \cdot cap(B,d)^e}
\vspace{-0.1in}
\end{equation}
where $c_d$ is the coefficient for the resource component $d$ and 
the exponent $e$ is a measure of cost sensitivity to the changes in the resource quantities.
We shall consider two cost models: homogeneous linear and heterogeneous. 
In the former, the coefficients and the exponent are set to one.
In the latter, the coefficients are set heterogeneously and the exponent is varied
to simulate non-linearity: 
(i) when $e=1$, the per-unit rate (cost/quantity) of a resource component remains constant;
(ii) when $e<1$, the rate decreases with increase in the quantity;
(iii) when $e>1$, the rate increases with increase in quantity.

\mypara{Algorithms} 
Taking the penalty based mapping ({\penmap}) as the baseline,
we evaluate the LP based mapping ({\LPmap}), LP based mapping with cross node-type filling ({\LPmapF}) and penalty based mapping with cross node-type filling ({\penmapF}).

For {\penmap} and {\penmapF}, we report the minimum cost obtained from four (i.e. two mapping and two fitting policies described in Section~\ref{sec:penmap}) combinations each. Similarly, for {\LPmap} and {\LPmapF}, we take the minimum cost from two fitting policies respectively.

\mypara{Linear Programming Lowerbound}
We use the linear program discussed earlier (Section~\ref{sec:lp-lb}) to derive a lowerbound ({\LB}) on the cost of the optimal solution.
The cost of solutions $S$ output by the algorithms are normalized as $\cost(S)/{\LB}$,
so that a normalized cost of $1$ means that $S$ is optimal. 

\subsection{Homogeneous Linear Cost-Model}
\label{sec:homcm}
As noted earlier, the homogeneous cost-model
computes $\cost(B)$ by setting $e=1$ and $c_d=1$ (Equation \ref{eq:cm}).

\begin{figure*}[!ht]
\centering
\begin{subfigure}[h]{0.32\textwidth}
    \centering
    \includegraphics[width=\textwidth]{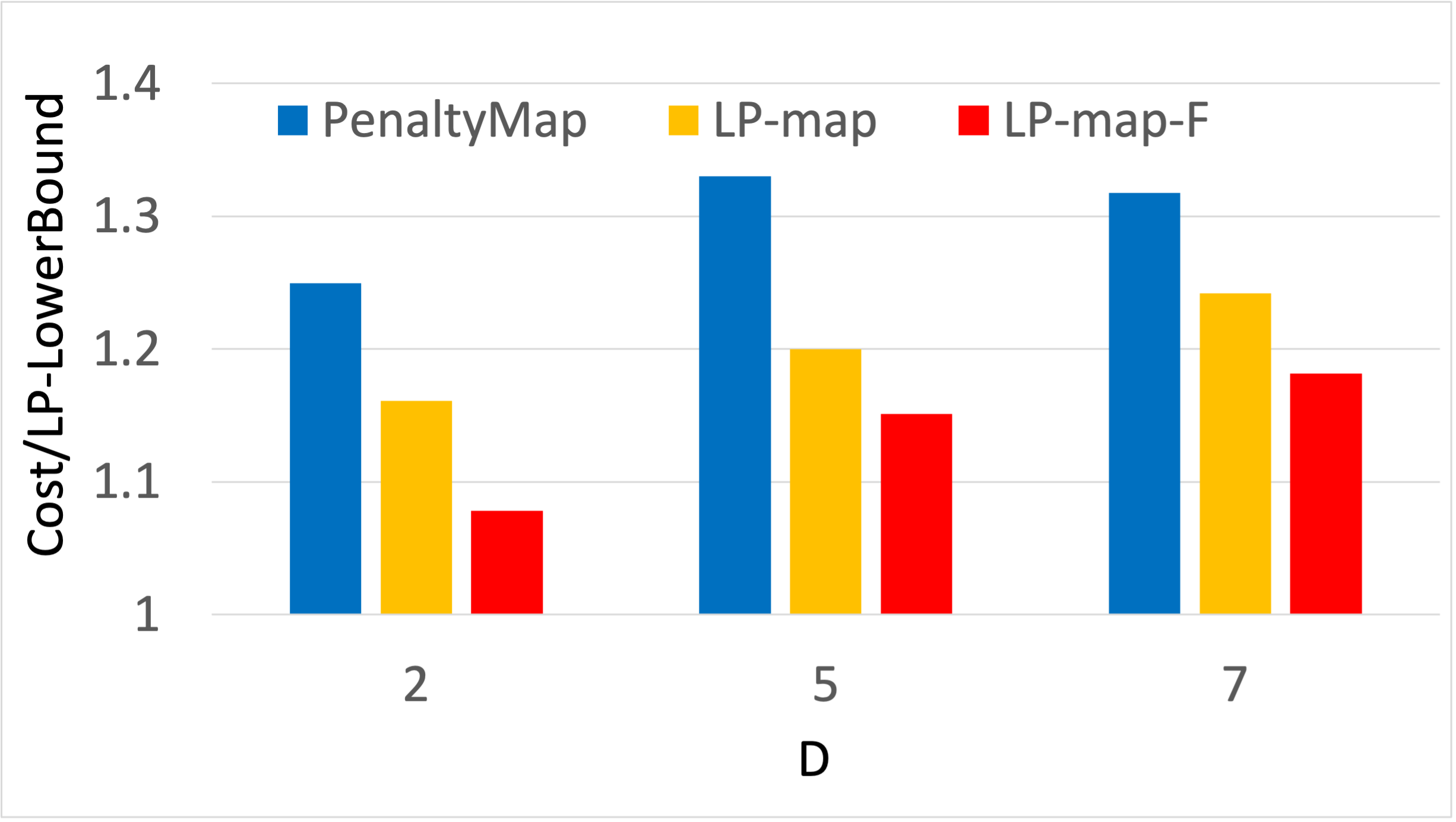}
    \caption{Varying $D$. $m$=$10$, demand=$[0.01, 0.1]$}
    \label{fig:chart1a}
\end{subfigure}
\begin{subfigure}[h]{0.32\textwidth}
    \centering
    \includegraphics[width=\textwidth]{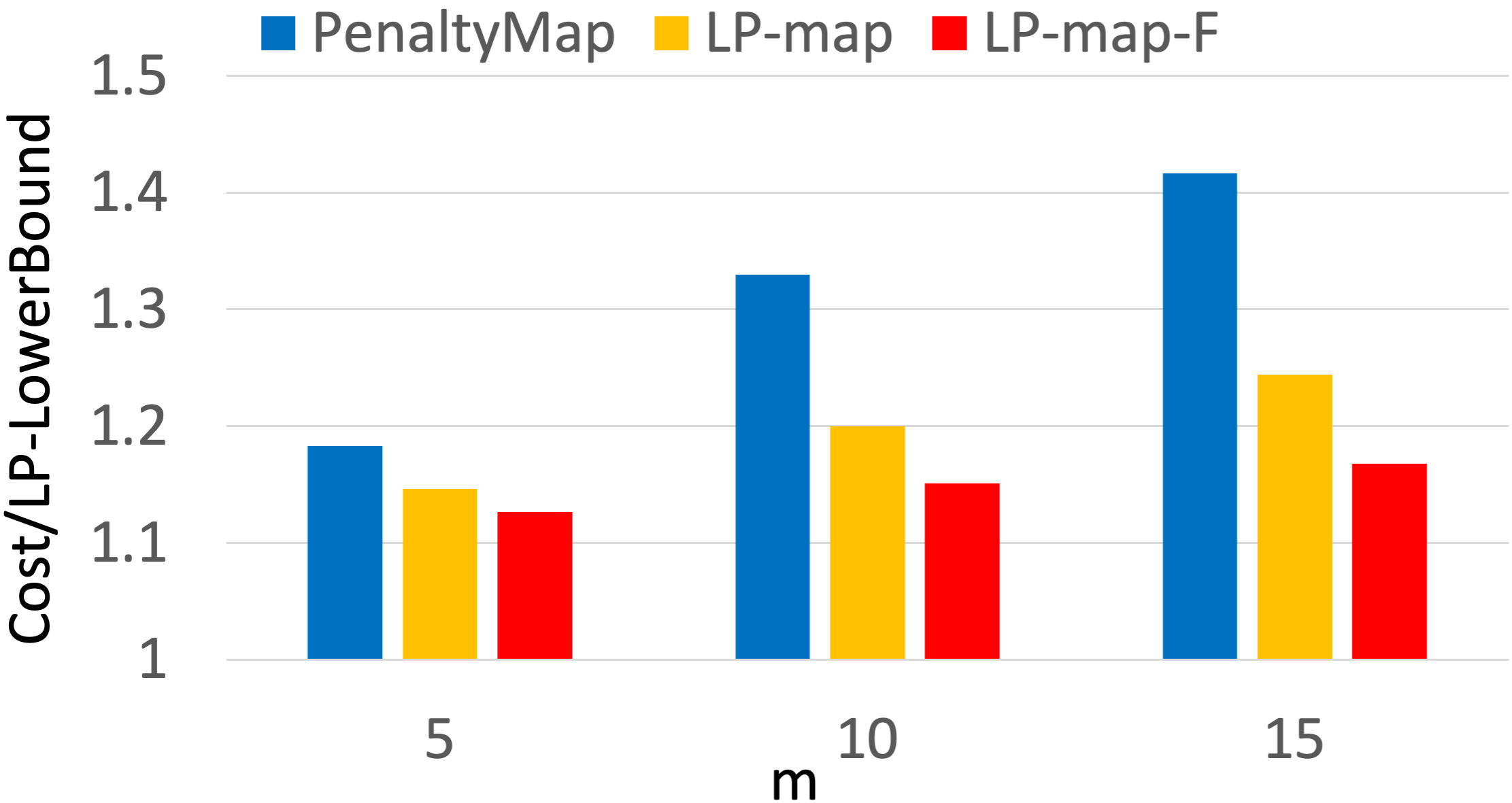}
    \caption{Varying $m$. D=$5$, demand=$[0.01, 0.1]$}
    \label{fig:chart1b}
\end{subfigure}  
\begin{subfigure}[h]{0.32\textwidth}
    \centering
    \includegraphics[width=\textwidth]{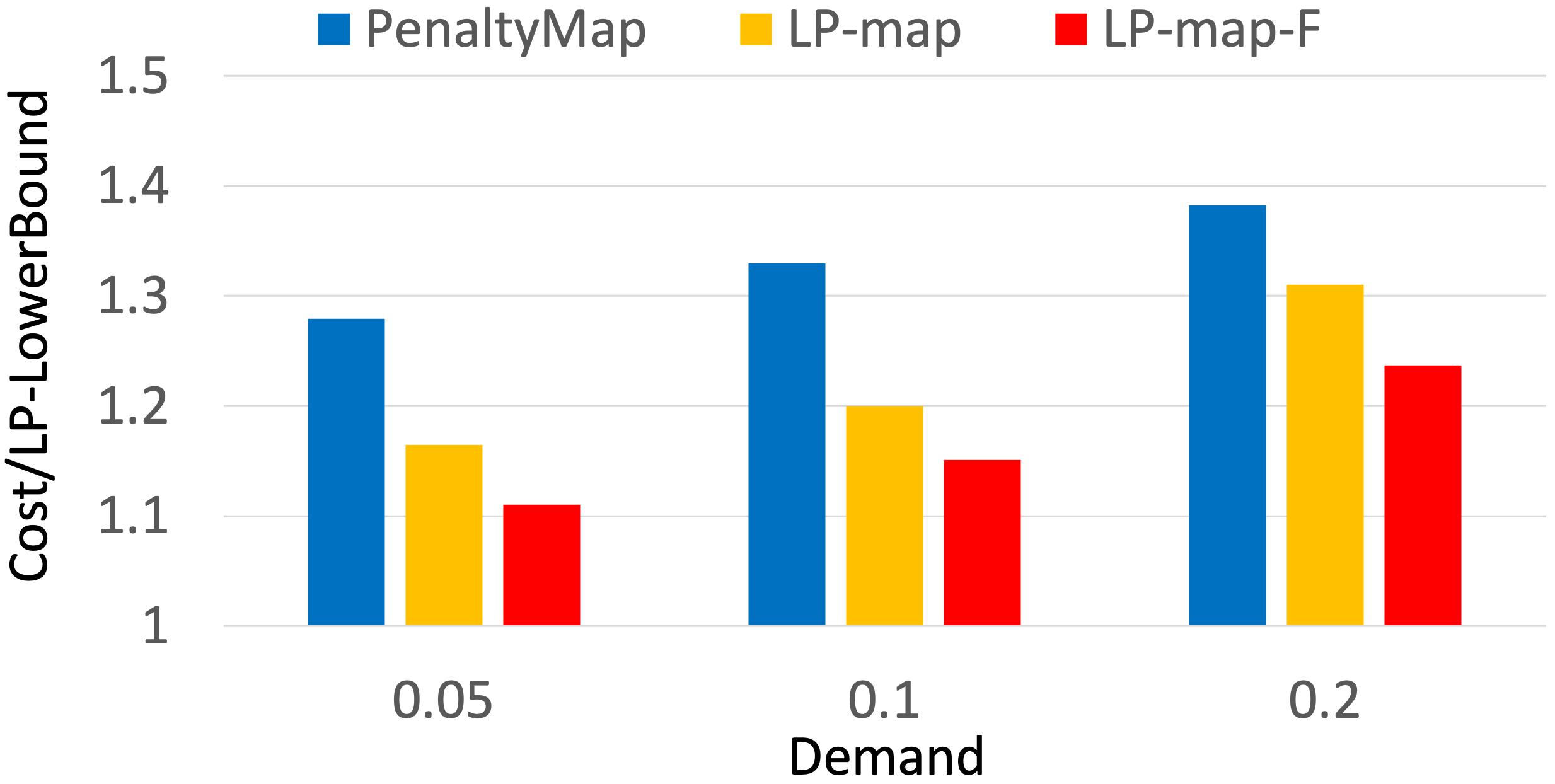}
    \caption{Varying demand. $m$=$10$, $D$=$5$}
    \label{fig:chart1c}
\end{subfigure}  
\caption{[Synthetic-Homogeneous]: Comparison of algorithm when $m$, $D$ and demand are scaled.}
\label{fig:chart1}
\vspace{-0.15in}
\end{figure*}
\subsubsection{Synthetic Trace}
We first compare the three algorithms by scaling each of $D$, $m$ and the demand,
fixing the other parameters.

\mypara{Dimension}
In Figure~\ref{fig:chart1a} $D$ is varied from $2$ to $7$.
Compared to {\penmap}, the {\LPmap} strategy decreases the normalized cost by up to $0.13$ (or $13\%$),
and {\LPmapF}  plugs wastage by using the cross node-type filling, resulting in further reduction by up to $8\%$,
taking the solution closer to the optimal.
Overall, {\LPmapF} offers up to $17\%$ decrease in cost compared to {\penmap}. 

We also note that cost increases for all the mapping strategies as $D$ increases, 
since the instances become harder, but {\LPmapF} remains more resilient and 
remains within 20\% of the lowerbound even for higher $D$,
and  is well within 10\% for lower $D$, indicating that solution is close to optimal.

\mypara{Node-type}
In Figure~\ref{fig:chart1b} $m$ is varied from 5 to 15. 
Here again, {\LPmapF} outperforms {\penmap} by up to 24\% with  cross node-type filling contributing up to 7\% in this cost reduction.
As a general trend, more node-types makes the mapping issue harder and 
the cost of all algorithms increase. 
The {\penmap} algorithm is $18\%$ away from the lowerbound at $m=5$, and shoots to $41\%$ at $m=15$.
In contrast, the {\LPmapF} algorithm remains stable resulting in only $4\%$ increase ($1.12$ to $1.16$)

\mypara{Task demand}
In Figure~\ref{fig:chart1c} task demand is varied from [0.01, 0.05] to [0.01, 0.2]. 
As in the two prior cases, {\LPmapF} performs the best in all the cases 
and stays under 25\% of the lowerbound. 

\subsubsection{Real-trace Performance $-$ Scaling Tasks and Node-types}
\begin{figure}[t]
\centering
\begin{subfigure}[h]{0.32\textwidth}
    \centering
    \includegraphics[width=\textwidth]{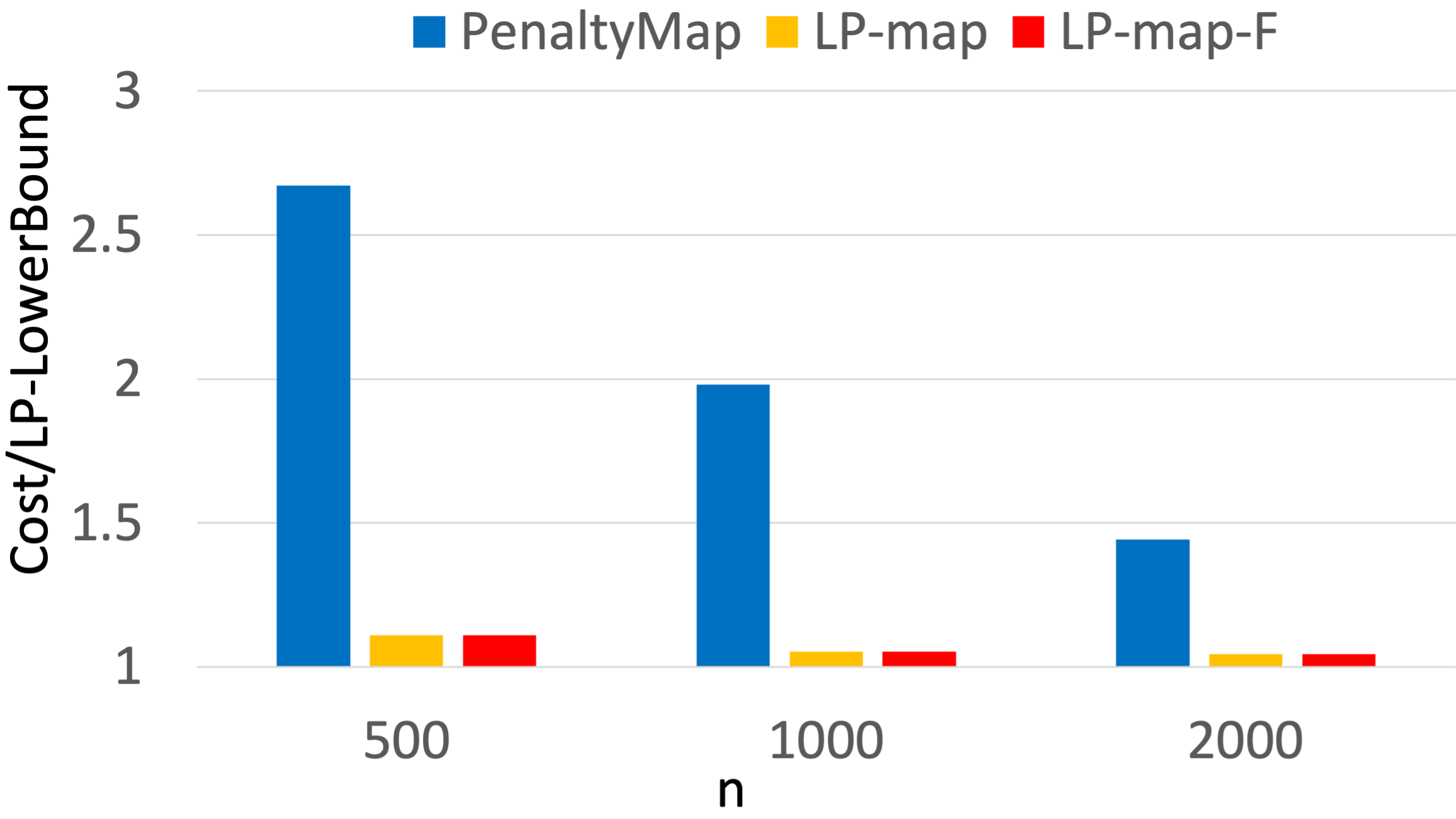}
    \caption{Varying $n$. $m$=$10$}
    \label{fig:chart2a}
\end{subfigure}
\begin{subfigure}[h]{0.32\textwidth}
    \centering
    \includegraphics[width=\textwidth]{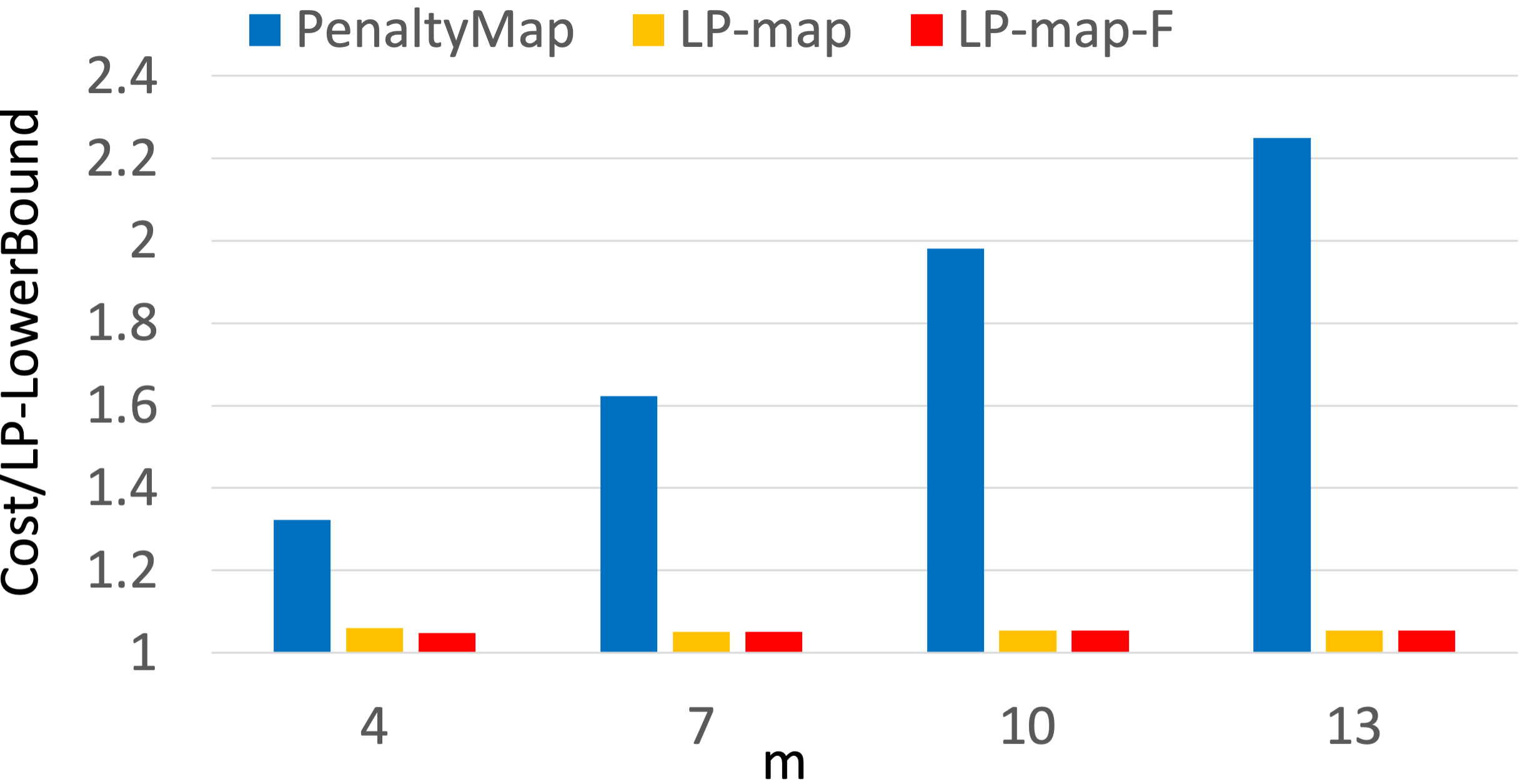}
    \caption{Varying $m$. $n$=$1000$}
    \label{fig:chart2b}
\end{subfigure}  
\caption{[{\google}-Homogeneous]: Quality of solution when $m$ and $n$ are scaled.}
\label{fig:chart2}
\vspace{-0.15in}
\end{figure}

Google trace results are shown in Figure~\ref{fig:chart2}.
First, we scale $n$ (Figure~\ref{fig:chart2a}) while keeping $m=10$ constant,
and $D=2$ (CPU and memory).
We see that {\LPmap} outperforms {\penmap} by large margins from $39\%$ to $156\%$
and produces near-optimal solutions, not more than $11\%$ away from the lowerbound.
Our second experiment (Figure \ref{fig:chart2b}) fixes $n=1000$ and varies $m$ from $4$ to $13$,
wherein {\LPmap} maintains $26\%-119\%$ lower cost than {\penmap}
and remains under $7\%$ of lowerbound. 

In {\google} trace, the task demands are small and so, the LP mappings are near-integral.
Further, at smaller $D$,
it is easier to fit the tasks, causing {\LPmap} solutions to be very close to the optimal.
The  cross node-type filling has little room for improvement and hence {\LPmapF} exhibits similar performance as {\LPmap}.

As $n$ increases, {\penmap} has more tasks to map, 
due to which sub-optimal mappings get averaged, 
resulting in lesser capacity wastage and improved performance.
However, as $m$ increases, the mapping issue becomes harder
and the performance of the greedy heuristic degrades.

\subsection{Heterogeneous Cost-Model}
\label{sec:hetcm}
\begin{figure}[t]
\centering
\includegraphics[width=.8\linewidth]{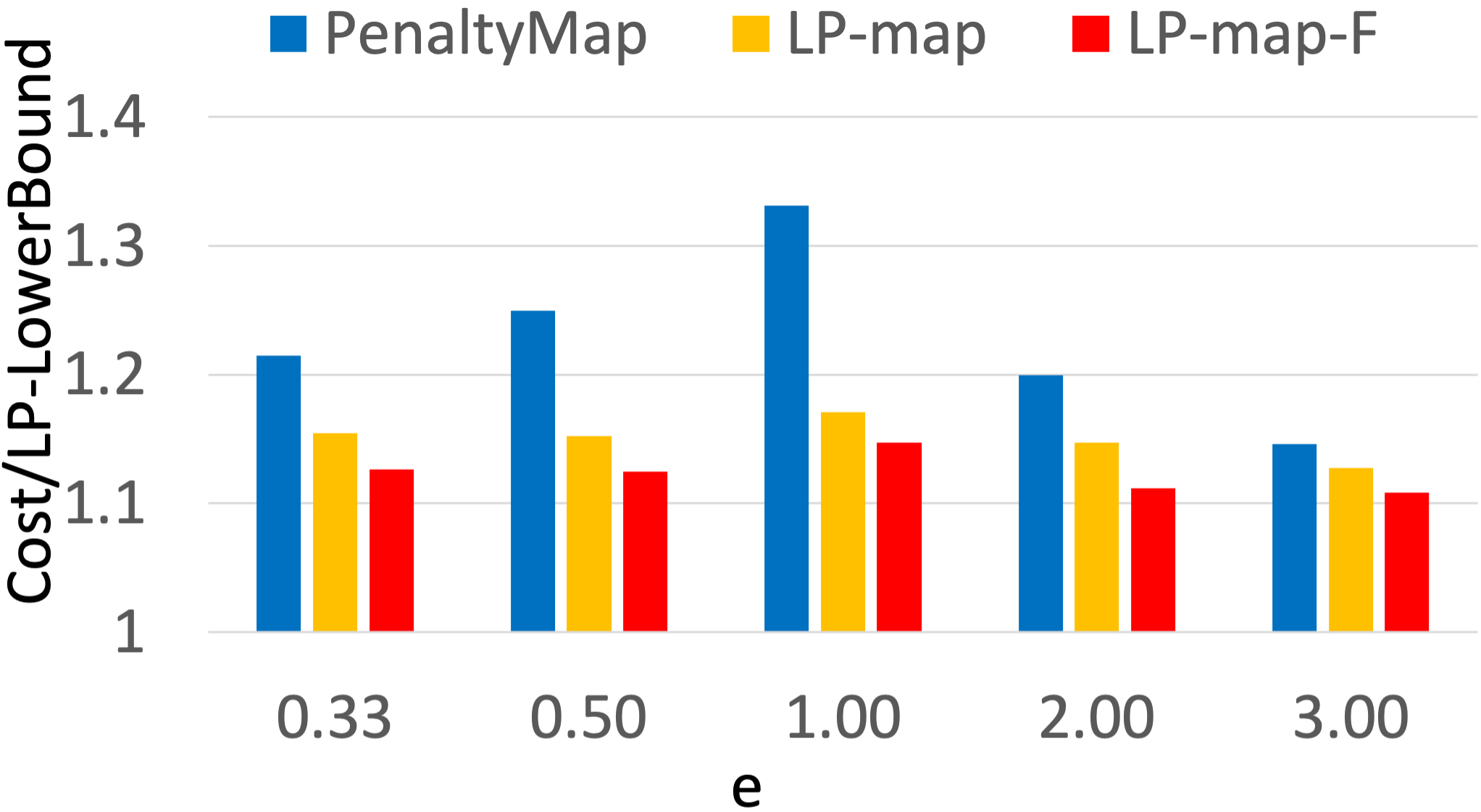}
\caption{[Synthetic-Heterogeneous]: Varying $e$ for a given $m$.}
\label{fig:chart3a}
\vspace{-0.1in}
\end{figure}

\mypara{Synthetic Data}
We use a median point of $D=5$, $m=10$ and demand = $[0.01, 0.1]$ for this experiment. Coefficients $c_d$ for the cost model (Equation \ref{eq:cm}) are generated randomly in the range of $[0.3, 1.0]$. Further, we vary $e$ from $0.33$ to $3$.
The results (Figure~\ref{fig:chart3a}) show that {\LPmapF} offers up to $18\%$ gain over {\penmap} for $e=1$, and up to $12\%$ gain $e < 1$.
Cross node-type filling provides a constant benefit of $2\%-3\%$ over {\LPmap} for all values of $e$.

When $e > 1$, the cost is skewed towards the largest resource component,
and when $e < 1$, the model tends towards uniform cost across the node-types (irrespective of their capacity).
In both the cases, it is easier to map the tasks and hence, the greedy {\penmap} gets closer to {\LPmap}.
 
\mypara{Real-life Trace {\google}} 
For each node-type in GCT-2019, we use pricing coefficients from the Google pricing model~\cite{gct2019pm}, keeping the exponent $e=1$.
\begin{figure}[t]
\centering
\includegraphics[width=.8\linewidth]{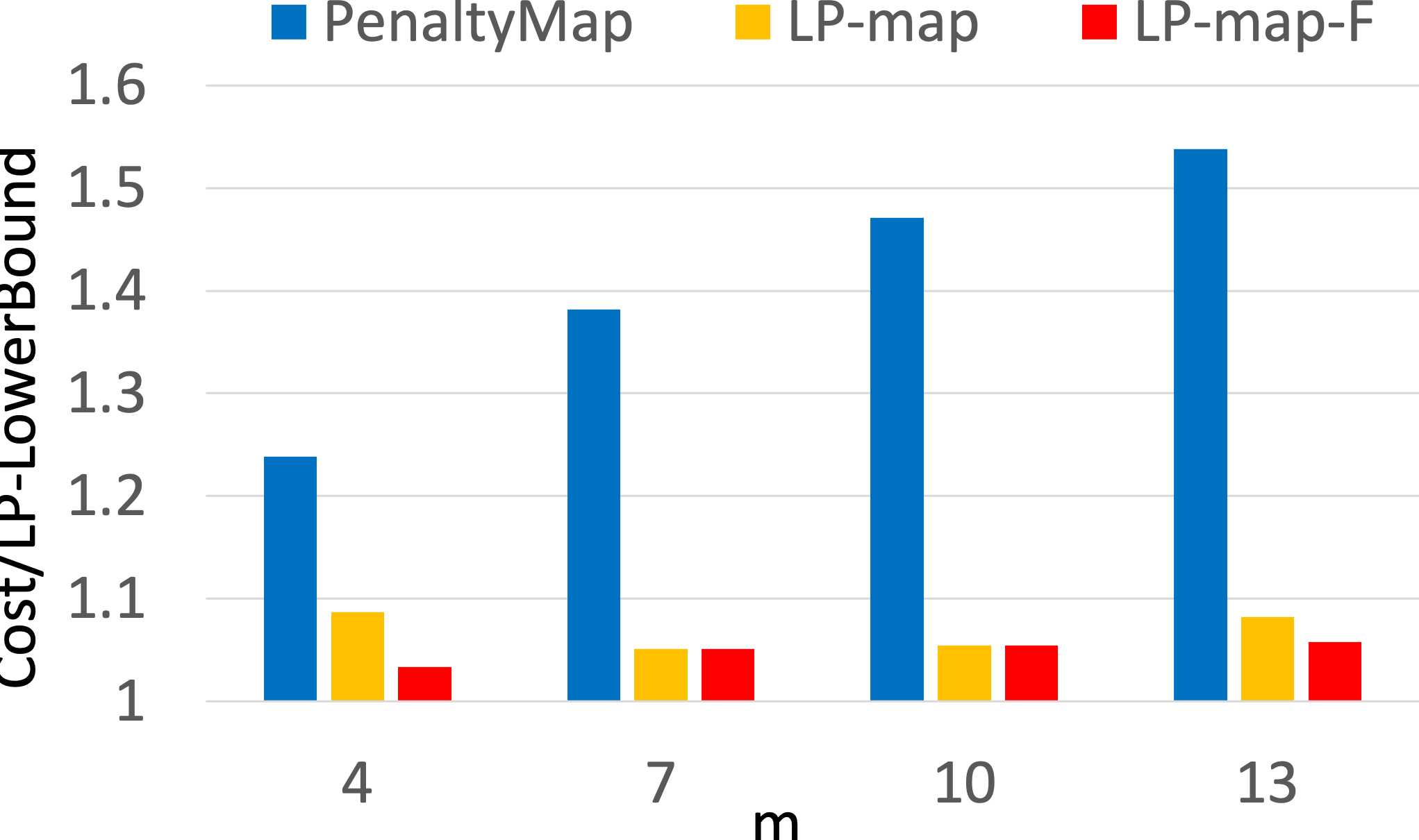}
\caption{[{\google}-Heterogeneous]: Varying node-types ($n$=$1000$, $e$=$1$) using cost model with real-world coefficients.}
\label{fig:chart3b}
\vspace{-0.2in}
\end{figure}
Since, number of node-types has an important role in the quality of solution, we vary the node-types from $4$ to $13$. 
The results are shown in Figure~\ref{fig:chart3b}. 
The performance of {\penmap} is away from the lowerbound by $23\%$ at $m=4$, and degrades as $m$ increases,
reaching $53\%$ at $m=13$.
In contrast, {\LPmapF} gives high quality solutions staying close to $5\%$ of the lowerbound, 
while outperforming {\penmap} by up to $48\%$. Cross node-type filling contributes up to $5\%$ in {\LPmapF} performance. 

\subsection{Cross Node-Type Filling}
As mentioned earlier (Section \ref{sec:cross}), cross node-type filling can be applied over {\penmap} as well,
and it is interesting to study the implications.
Towards that goal, we augment {\penmap} with the filling method to derive a {\penmapF} version,
and evaluate on all the data points of the {\google} trace experiments (Figures \ref{fig:chart2} and \ref{fig:chart3b}).
We can see that the filling method offers $44\%$-$142\%$ gains over {\penmap}.
\begin{figure}[t]
\centering
\includegraphics[width=.9\linewidth]{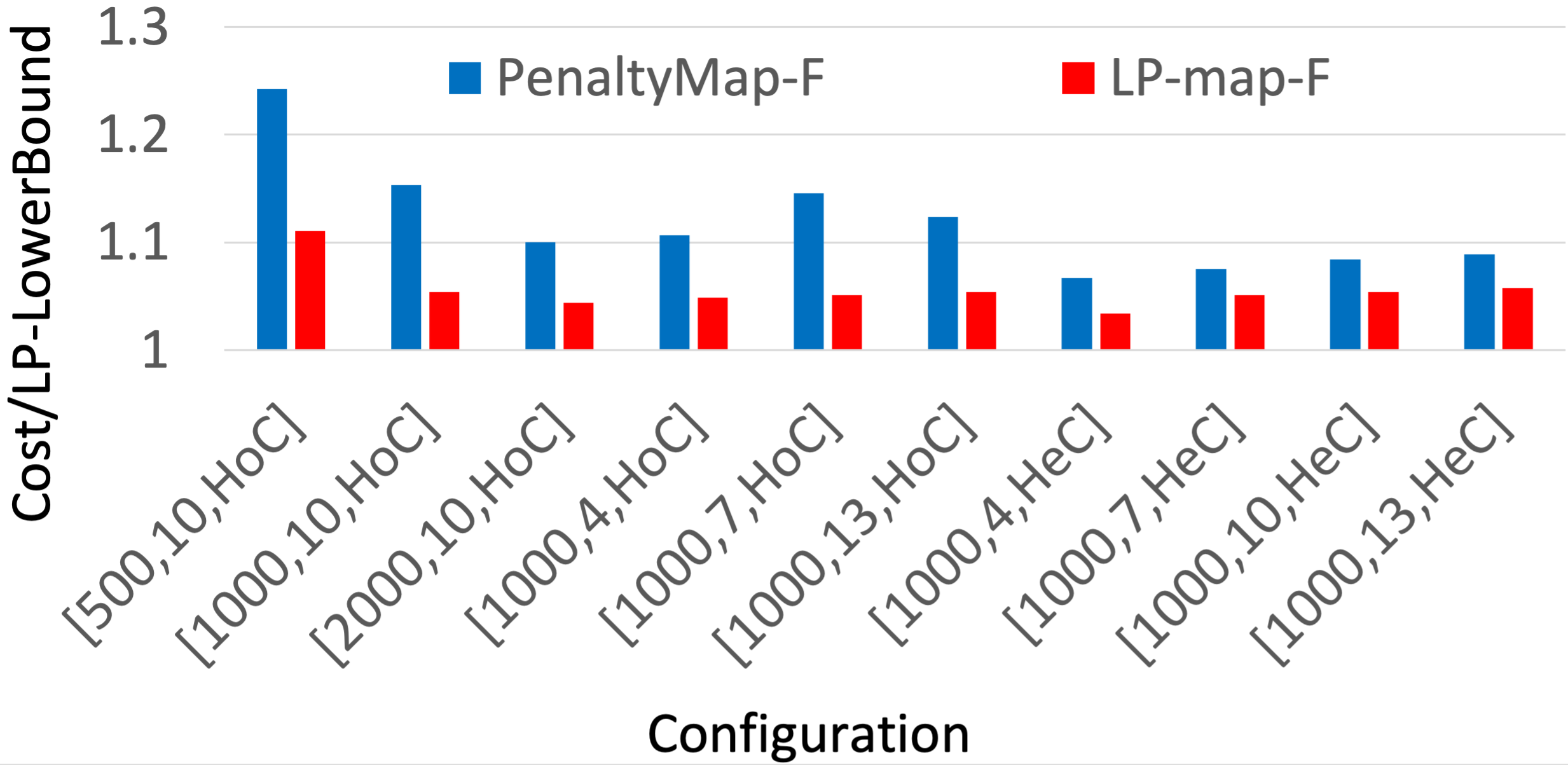}
\caption{[{\google} All-Scenarios]: Comparing {\penmapF} and {\LPmapF}.}
\label{fig:chart4}
\vspace{-0.25in}
\end{figure}
Figure~\ref{fig:chart4} compares {\penmapF} and {\LPmapF}, where {\LPmapF} outperforms {\penmapF} 
consistently with gains in the range $3\%$-$13\%$ establishing that an organic combination of LP-based mapping and 
cross node-type filling provides better solutions.

\subsection{Running Time}
We profiled the running time for all the algorithms on the largest configuration of 2000 tasks and 
13 node-types from the {\google} trace.
We find that on the average, {\penmap} takes about one second to produce a solution. 
Running time of the LP-based mappings could be split 
into the running time of the LP solver and the mapping phase. While the LP solver itself takes about 15 min, 
the mapping phase  for both {\LPmap} and {\LPmapF} takes about a second. 
Since our objective is cold-starting the cluster and cluster-sizing 
is an one-time computation, 15min is a very practical running time.

\subsection{No-Timeline Scenario} 
Figure~\ref{fig:motivate} illustrates the gains of incorporating the timeline into solution computation. 
To experimentally confirm these gains, we considered the configuration of 1000 tasks and 10 node-types from the {\google} trace.
We compared the cluster cost under timeline-aware and timeline-agnostic settings.
For the former, we use the cost of {\LPmapF} solution (Figure \ref{fig:chart2a}) and
for the latter, we compute a lowerbound on the cost by treating the tasks to be always active.
As expected, the latter produces higher cost, and  we found that the factor is as high as $2$x on the average.

\eat{
- PEN runs in time within 10 seconds on all - where it reached.
- LP solving -- max time solver.
- Given LP solution, lp and lpf ran withi 10 seconds.
}

\eat{
It is obvious that time conscious is better. We confirm ... .
}

\section{Conclusions}
We introduced the {\TLrightsize} problem that reduces cluster cost by exploiting the time-limited nature of real-life tasks. 
We designed a baseline two-phase algorithm with an approximation ratio of $O(D\cdot \min(m, T))$,  
an improved LP-based mapping strategy and a cross-node-type filling method for further cost reduction.
Experiments on synthetic and real-life traces demonstrate that 
the final  algorithm produces solutions
within $20\%$ of a lowerbound on the optimal cost. 
We identify two avenues of fruitful research. First is to further bridge the gap between the lowerbound
and the solutions on the hard instances. Secondly, improving the approximation 
ratio would be of interest from  a theoretical perspective.
We also intend to work on enhancing the scheduler and auto-scaling algorithms to better leverage the output from {\TLrightsize}

\bibliographystyle{plain}
\bibliography{main}
\end{document}